\documentclass[twocolumn]{aa}

\usepackage{graphicx}
\usepackage{txfonts}
\begin{document}
\def\lsim{\,\lower2truept\hbox{${< \atop\hbox{\raise4truept\hbox{$\sim$}}}$}\,}
\def\gsim{\,\lower2truept\hbox{${> \atop\hbox{\raise4truept\hbox{$\sim$}}}$}\,}

   \title{A multifrequency angular power spectrum analysis of the 
Leiden polarization surveys}

   \author{L.~La~Porta\inst{1,}\thanks{Member of the 
          International Max Planck Research Shool (IMPRS) for
	  Radio and Infrared Astronomy at the Universities
	   of Bonn and Cologne.}          
	  \and
          C.~Burigana\inst{2}
          }

   \offprints{C.~Burigana}

   \institute{Max Planck Institut f\"ur Radioastronomie,
              Auf dem H\"ugel, 69, D-53121 Bonn, Germany\\
              \email{~laporta@mpifr-bonn.mpg.de}
             \and 
              INAF-IASF Bologna, 
              via P.~Gobetti, 101, I-40129 Bologna, Italy\\
              \email{~burigana@iasfbo.inaf.it} \\
             }

   \date{Sent October 6, 2005; accepted May 15, 2006}

   \abstract{ The Galactic synchrotron emission 
is expected to be the most relevant source of astrophysical contamination
 in cosmic microwave background polarization measurements, at least 
at frequencies $\nu \lsim 70$~GHz and at angular scales $\theta\gsim 30'$.  
We present a multifrequency analysis of the Leiden surveys, 
linear polarization surveys covering the 
Northern Celestial Hemisphere at five frequencies between 408~MHz and 
1411~MHz. By implementing specific interpolation methods to 
deal with these irregularly sampled data, we produced maps of the 
polarized diffuse Galactic radio emission with a pixel size 
$\simeq 0.92^\circ$. We derived the angular power spectrum (APS)
($PI$, $E$, and $B$ modes)
of the synchrotron dominated radio emission
as function of the multipole, $\ell$.
We considered the whole covered region and 
some patches
at different Galactic latitudes. 
By fitting the APS in terms of power 
laws ($C_\ell \sim \kappa\cdot\ell^{\alpha}$),
we found spectral indices that steepen with increasing frequency:
from $\alpha \sim -$(1-1.5) at 408~MHz 
to $\alpha \sim -$(2-3) at 1411~MHz 
for $10 \lsim \ell \lsim 100$ and from $\alpha \sim -0.7$ 
to $\alpha \sim -1.5$ for lower multipoles 
(the exact values depending on the considered sky region 
and polarization mode). The bulk of this flattening
at lower frequencies can 
be interpreted in terms of Faraday depolarization effects.
We then considered the APS at various fixed multipoles and
 its frequency dependence. Using the APSs of the Leiden surveys
 at 820~MHz and 1411~MHz, we determined possible ranges
for the rotation measure, $RM$, in the simple 
case of an interstellar medium {\it slab model}.
Also taking into account the polarization degree at 
1.4~GHz, 
it is possible to break the degeneracy between the
identified $RM$ intervals.
The most reasonable of them turned out to be 
$RM \sim 9-17$~rad/m$^2$
although, given the uncertainty on the measured polarization
degree, $RM$ values in the interval $\sim 53-59$~rad/m$^2$
cannot be excluded.

   }

\authorrunning{L.~La~Porta \& C.~Burigana}
\titlerunning{Multifrequency APS analysis of Leiden surveys}

   \maketitle
%

\section{Introduction}

In the last decade an impressive number of 
experiments has been dedicated to measurements of the 
cosmic microwave background 
(CMB) anisotropies, first
observed in temperature by COBE (\cite{smoot1990}). 
Modern cosmologies predict also the existence of
polarization fluctuations, produced at the recombination epoch via Thomson 
scattering (see, e.g., \cite{Kosowsky99}). 
However, the foreseen degree of polarization of the CMB anisotropies should  
be $\lsim 10\%$, implying a signal much weaker than in temperature and 
therefore extremely difficult to reveal. 
The first detection of the CMB polarization
achieved by DASI (\cite{Kovac2002}) and the recent 
measure by BOOMERanG (\cite{masietal05};~\cite{montroyetal05}) 
 confirmed the expectations. The polarization data 
expected by the NASA WMAP~\footnote{http://lambda.gsfc.nasa.gov/product/map/} 
satellite will permit us to extend the information about polarization 
to a wider portion of the sky, though at small scales ($\theta \lsim 0.5^{\circ}$)
its sensitivity will likely allow a detection rather than an accurate measure.
 
In the near future, the ESA 
{\sc Planck}~\footnote{http://www.rssd.esa.int/planck}  satellite will 
significantly improve these results by observing the whole sky
at nine different frequencies between 30 and 857~GHz, both in temperature 
and polarization, with unprecedented resolution and sensitivity (e.g. see  
\cite{tauber04} and references therein). 
In particular, the sensitivities per pixel of the {\sc Planck} maps are 
expected to be between 3-4 and 10 times better than in the WMAP data, 
producing a significant improvement in terms of angular power spectrum 
recovery.
Consequently, {\sc Planck} alone will allow an estimate of the cosmological 
parameters much more precise than that obtained by WMAP and other CMB 
experiments combined with different cosmological data. 
Moreover, it will accurately measure the CMB $E$-mode and could 
potentially observe the $B$-mode as well 
(e.g. see \cite{burigana04} and references therein).
 
Strong efforts have been made to evaluate the impact of the foregrounds 
on the feasibility of CMB anisotropy measurements, as well as to
understand how to separate the different components contributing to the 
observed maps. 
The methods elaborated for this purpose can be divided in two
categories: blind (Independent Component Analysis, see \cite{ica};
\cite{fastica}; \cite{icapol}; \cite{delabrouille})
and non-blind (Wiener filtering and Maximum Entropy Methods, 
see respectively \cite{wf} and \cite{mem}). 
The latter class needs an a priori
knowledge of the frequency and spatial dependence
of the foreground properties, 
while a blind approach only requires maps at different 
frequencies as input. On the other hand, blind methods  
benefit from having ancillary data (such as realistic templates) 
to improve the separation robustness and 
quality~\footnote{Future component separation 
methods will probably 
search for a good compromise between the 
relaxation of the a-priori assumptions on the independent signals 
superimposed on the microwave sky and the exploitation of the physical
correlation among the various components. Also 
it will be crucial to analyse microwave data together with
ancillary data from radio and infrared frequencies.}.

The Galactic polarized diffuse synchrotron radiation is expected
to play a major role at frequencies below 70~GHz on intermediate
and large angular scales ($\theta \gsim 30'$), 
at least at medium and high Galactic latitudes where satellites have the 
clearest view of the CMB anisotropies.
At about 1~GHz the synchrotron emission is the most important radiative
mechanism out of the Galactic plane, while at low latitudes it is 
comparable with the bremsstrahlung; 
however the free-free emission is 
unpolarized, whereas the 
synchrotron radiation could reach a theoretical intrinsic
degree of polarization of about $75\%$. 
Consequently, radio frequencies are the natural range for studying the
Galactic synchrotron emission, 
although it might be affected by Faraday rotation and depolarization.
Nowadays\footnote{During the final phase of the revision process
of this work, the DRAO 1.4 GHz polarization survey has been released
(\cite{wolleben05}). The data are available at 
http://www.mpifr-bonn.mpg.de/div/konti/26msurvey/ or
http://www.drao-ofr.hia-iha.nrc-cnrc.gc.ca/26msurvey/ or 
http://cdsweb.u-strasbg.fr/cgi-bin/qcat/?J/A+A/448/411~.
See La Porta et al.~2006 for a first angular power
spectrum analysis of the DRAO survey.} 
the only available data suitable for studying the polarized Galactic 
synchrotron emission on large scales are the so-called Leiden surveys
(\cite{spo76}). 
These are linear polarization surveys, extending up to high 
Galactic latitudes, carried out at five frequencies 
between 408~MHz and 1411~MHz. 
We have elaborated specific interpolation methods 
to project these surveys into maps 
with a pixel size $\simeq 0.92^{\circ}$.
These maps can be exploited in different contexts.
They are suitable to analyse the statistical
properties of the Galactic radio polarized emission.
Beside this obvious application, they can be used as input 
to construct templates for simulation activities in the 
context of current and future microwave polarization 
anisotropy experiments. 
For example they can be utilized to build templates 
for straylight evaluation (see, e.g., 
\cite{challinoretal2000};~\cite{buriganaetal2001};
~\cite{barnesetal2003}) and for component separation analyses.

It is standard practice to characterize the CMB anisotropies  
in terms of the angular power spectrum~\footnote{It contains
all the relevant statistical information 
in the case of pure Gaussian fluctuations.} (APS)
as a function of the multipole, $\ell$ (inversely proportional to 
the angular scale, $\ell \simeq {180}/{\theta(^{\circ})}$).
The APS is an estimator related to the angular two-point correlation 
function of the fluctuations of a sky field (\cite{peebles}).
The APS is not able to fully characterize the complexity of the 
intrinsically non-Gaussian Galactic emission.
Nevertheless, it is of increasing use also in the study of the 
correlation properties of the Galactic foregrounds
(see, e.g., \cite{bacci01} for an application 
to the polarized radio emission).
We stick to the commonly adopted approach and 
consider the APS of both the polarized intensity, $PI$, 
and the $E$ and $B$ modes~\footnote{We keep here the physical dimension 
of the maps (antenna temperature, typically in K or mK) 
and consequently express the APS in K$^2$ or 
mK$^2$.} (see \cite{kamion} and \cite{zald}).

In Sect.~2 we briefly summarize the properties of the Leiden surveys
and of some selected areas relevant for the analysis at the smaller
angular scales. In Sect.~3 we introduce the general 
criteria adopted to elaborate the class of our interpolation algorithms.
The final, optimized version of the algorithm and the accuracy 
of the derived maps and APSs are described in 
Sect.~4. In Sect.~5 we present our results in terms of maps, APSs, and a 
parametric description of the APSs. A brief comparison with a preliminary
version of the DRAO polarization survey (\cite{wolleben03}; \cite{wollebenPhD}) at 
1.4~GHz is given in Sect.~6 in terms of APS. 
Sect.~7 is devoted to the multifrequency analysis 
of our results and to their interpretation in the context of an
 interstellar medium {\it slab model}.
We summarize and discuss our main results in Sect.~8.

\section{The Leiden surveys and selected areas}

The so-called Leiden surveys are the final results
of different observational campaigns carried out in 
the sixties with the 25-m radio telescope at Dwingeloo.
The complete data sets
as well as the observation reduction and calibration
methods have been presented by Brouw \& Spoelstra (1976). 
The observations were performed 
at 408, 465, 610, 820 and 1411~MHz with an angular resolution
of $\theta_{HPBW} = 2.3^{\circ}, 2.0^{\circ}, 
1.5^{\circ}, 1.0^{\circ}, 0.6^{\circ}$, respectively. 
At each frequency, the observed positions are inhomogenously
spread over the Northern Galactic Hemisphere 
and globally cover a sky fraction of about $\sim 50 \%$.
The errors, quoted for points observed at least twice, 
follow Gaussian distributions with mean values of 
$0.34, 0.33, 0.16, 0.11$ and $0.06$~K from
the lower to the higher frequency, respectively. 
The observations at the same position were taken at different 
azimuths, elevations, and days (the adopted scanning strategy being  
random in coordinates and time). 
A robust control and removal of the ground radiation contaminating
the observations have been also performed.
These properties imply a very low level of contamination from 
residual systematic effects, at least in comparison with the
intrinsic survey sensitivities, and make these surveys a 
reference point (though insufficient by itself) for absolute 
calibration of radio data. As an example, the absolute calibration
of the Effelsberg maps at 1.4~GHz  has been performed using 
the Leiden survey at 1411~MHz (see \cite{uya98}
for a description of the procedure).

The main problem in the analysis of the Leiden surveys is their 
poor sampling across the sky; it is necessary to project them 
onto maps with pixel size of about $3.7^{\circ}$
(we use here the HEALPix scheme, see 
\cite{gorski05}) to find at least one observation on each pixel of the  
observed region, that would limit the angular power spectrum
recovery to $\ell \simeq 50$. 
On the other hand, for some sky areas the sampling 
is significantly better than the average, by a factor $\simeq 4$, and the 
maps appear filled even for a pixel size of $\sim 1.8^{\circ}$, 
allowing one to reach multipoles $\ell \simeq 100$.
We identified three regions that permit one to investigate 
this multipole range at both low and middle/high Galactic latitudes:
patch 1 [($110^\circ \le l \le 160^\circ$, $0^\circ \le b \le 20^\circ$)];
patch 2 [($5^\circ \le l \le 80^\circ$, $b \ge 50^\circ$) together with
($0^\circ \le l \le 5^\circ$, $b \ge 60^\circ$) and
($335^\circ \le l \le 360^\circ$, $b \ge 60^\circ$)];
patch 3 [($10^\circ \le l \le 80^\circ$, $b \ge 70^\circ$)].

We underline that in these patches the polarized signal 
and the corresponding signal-to-noise ratio are typically higher 
than the average; they are associated with the brightest structures
of the polarized radio sky, i.e. the ``Fan Region'' (patch 1) and the
``North Polar Spur'' (patch 2 and 3). The ``North Polar Spur'' (NPS)
has been extensively studied (see \cite{salter83} 
and \cite{egger95}). 
The theories that better match the observations 
are all variants of the same model and interpret the 
NPS as the front shock of an evolved Supernova Remnant (SNR), 
whose distance should be $\sim 100\pm20$~pc
(inferred from starlight polarization, see \cite{bingham67}). 
In contrast, the present knowledge of the ``Fan Region'' 
is much poorer.
A rough estimate of its distance can be derived from 
purely geometrical considerations: it has such an extent on the sky
that it must be located within (1-2)$\times 10^2$~pc
from the Sun not to have an unrealistically large dimension.
Wolleben et al.~(2005) has recently proposed a different view
of the ``Fan Region'': from the distances of HII regions that they
consider to be located in front of this structure and
to depolarize its synchrotron emission, 
 they claim that the ``Fan Region'' is indeed an enormous
object extending up to the Perseus Arm, 
i.e. up to $\approx$~kpc distance.\\
Another common and remarkable characteristic of the selected
areas is that the polarization vectors appear mostly aligned 
at the two higher frequencies of the Leiden surveys.

\section{Map production algorithm and consistency tests}

A simple method to project the Leiden surveys into HEALPix
maps is to average the observations falling in each pixel.
However, the maps of the polarized intensity ($PI$) and of the 
Stokes parameters ($Q$,$U$) produced in this way show discontinuities
on scales $\theta \simeq \theta_{pixel}$. These discontinuities 
tend to add spurious power 
in the recovered APS, particularly at multipoles 
$\ell \sim 180/\theta_{pix}$.
In order to smooth these discontinuities 
 we implemented a ``class'' of specific 
``interpolation'' algorithms to generate maps with
$\theta_{pix}\sim 0.92^\circ$,  for the whole sky region
covered by the surveys.
To each pixel is assigned a weighted average of the signal values
falling in its neighbourhood, i.e.  
 $\sum_{i} ( p_{i} \cdot x_{i} )/ \sum p_{i}$
with $p_{i}=1/(\sigma_i^2 \cdot d_i^{n})$, 
where $x_i$ and $\sigma_i$
are the signal and the error associated with the observation
and $d_i$ its angular distance from the pixel centre.
The radius within which the average is computed varies  
pixel by pixel and is selected according to the following guidelines: 
(i) to have enough observations (at least 3);
(ii) to use only observations quite close to the
considered pixel centre (less than a few degrees);
(iii) to minimize the resulting average 
fractional change of the signal with the variation of the 
number of points and of the radius. 
The first two conditions imply that the interpolation
is as local as possible and is performed using a reasonable number
of data points. The third recipe is a convergency criterion.

In order to check the reliability of the method, we
generated at each frequency several groups of $PI$, $Q$, and $U$ 
maps, each corresponding to a different choice of the distance
power in the average weight ($1/(\sigma^{2}\cdot d^{n})$) for 
various  pixel sizes 
(ranging between $3.7^{\circ}$ and $0.92^{\circ}$).
A detailed description of the consistency tests together
with preliminary results referring to a first version
of the polarized intensity maps have been reported in
La~Porta~(2001) and Burigana \& La~Porta~(2002).
In that work we produced
simulated maps of white noise, expected to
contamine the astrophysical signal contained in the maps,
according to three different recipes.
We compared the APS of the simulated noise and signal maps
to identify the ranges of multipoles
statistically relevant
(i.e. where the signal is not masked by the noise)
for various considered sky coverages.
We found that the analysis of the survey full coverage
provides a good estimate of the polarization APS
for $\ell \sim [2,50]$, while the better sampled
patches (characterized by a higher S/N ratio on average)
allow us to investigate the multipole interval
$\sim [30,100]$.
We emphasize that the current APS analysis has a statistical
meaning and cannot provide details on the local properties
of the synchrotron emission.

\section{Interpolation algorithm optimization and accuracy}

\begin{figure*}  
   \centering
   \begin{tabular}{cc}
   \includegraphics[width=5cm,angle=90,clip=]{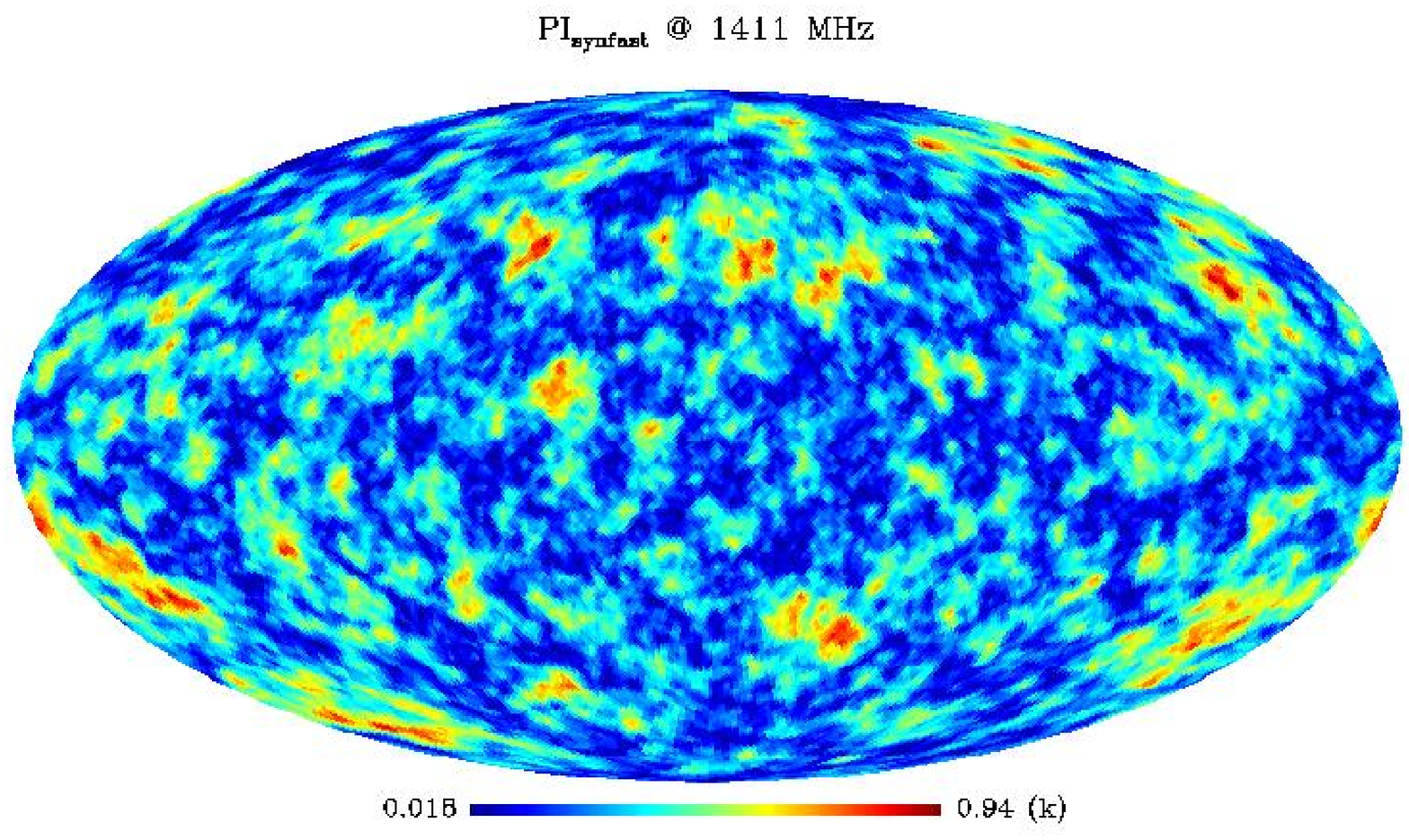}&
   \includegraphics[width=5cm,angle=90,clip=]{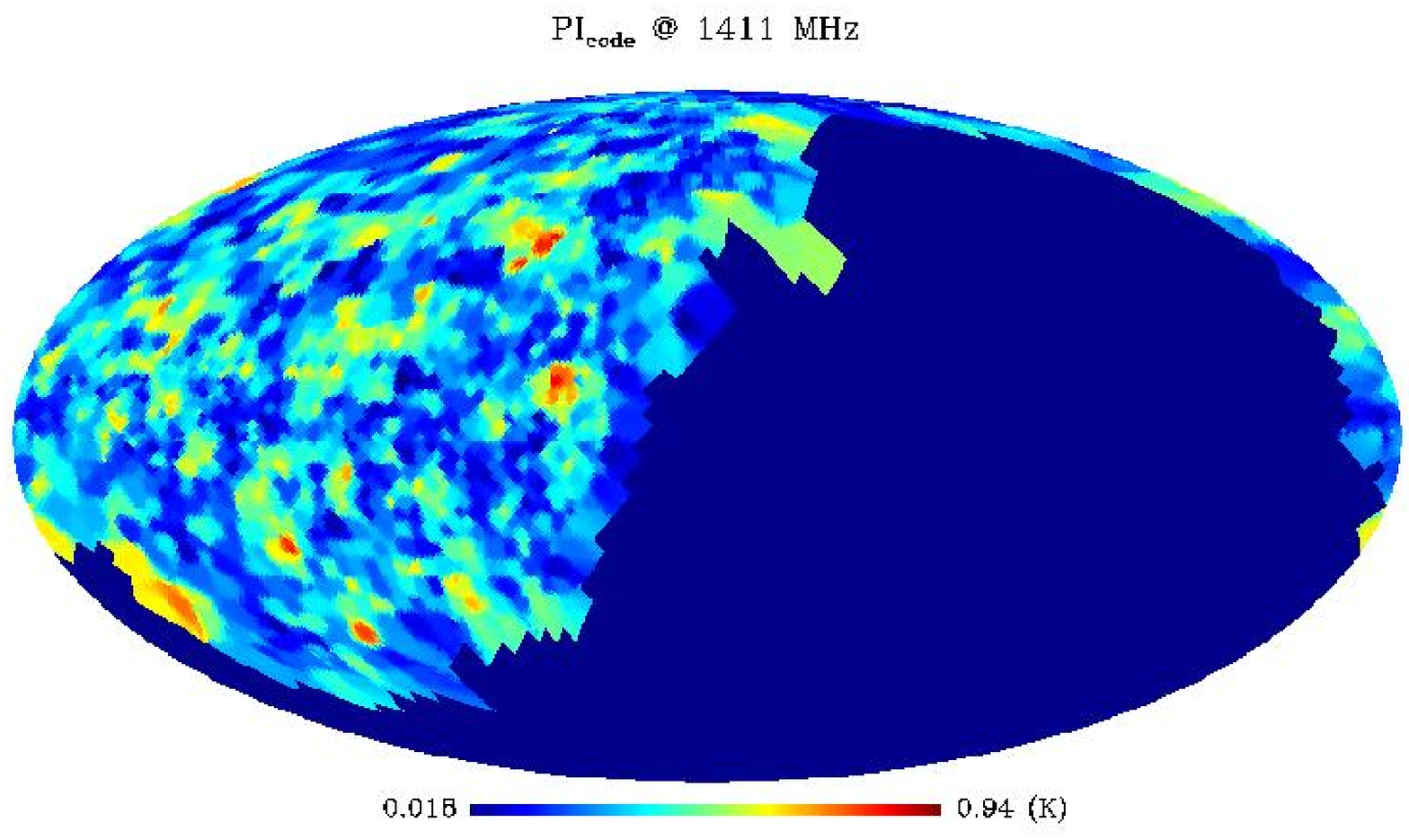}\\
   \includegraphics[width=5cm,angle=90,clip=]{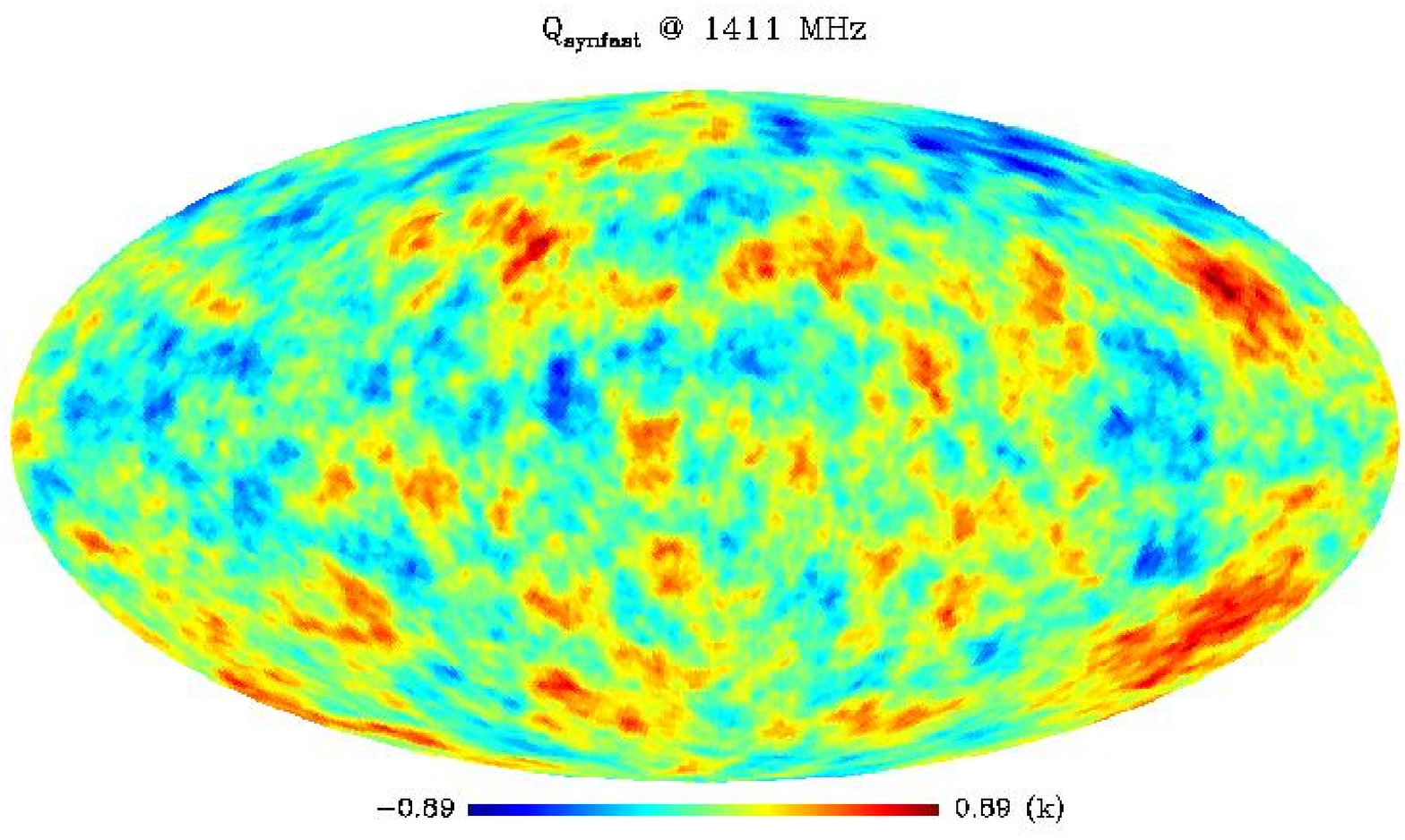}&
   \includegraphics[width=5cm,angle=90,clip=]{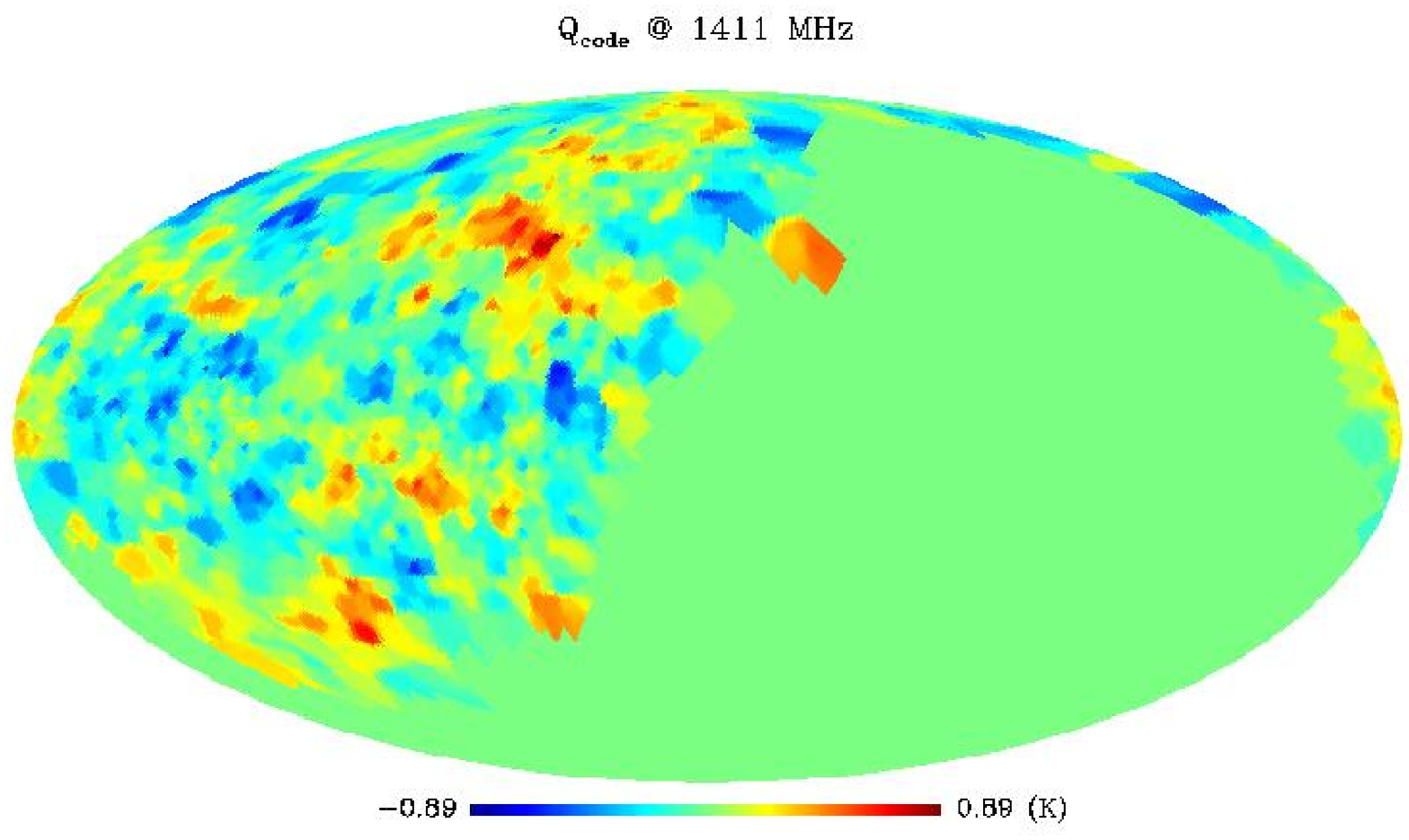}\\
   \includegraphics[width=5cm,angle=90,clip=]{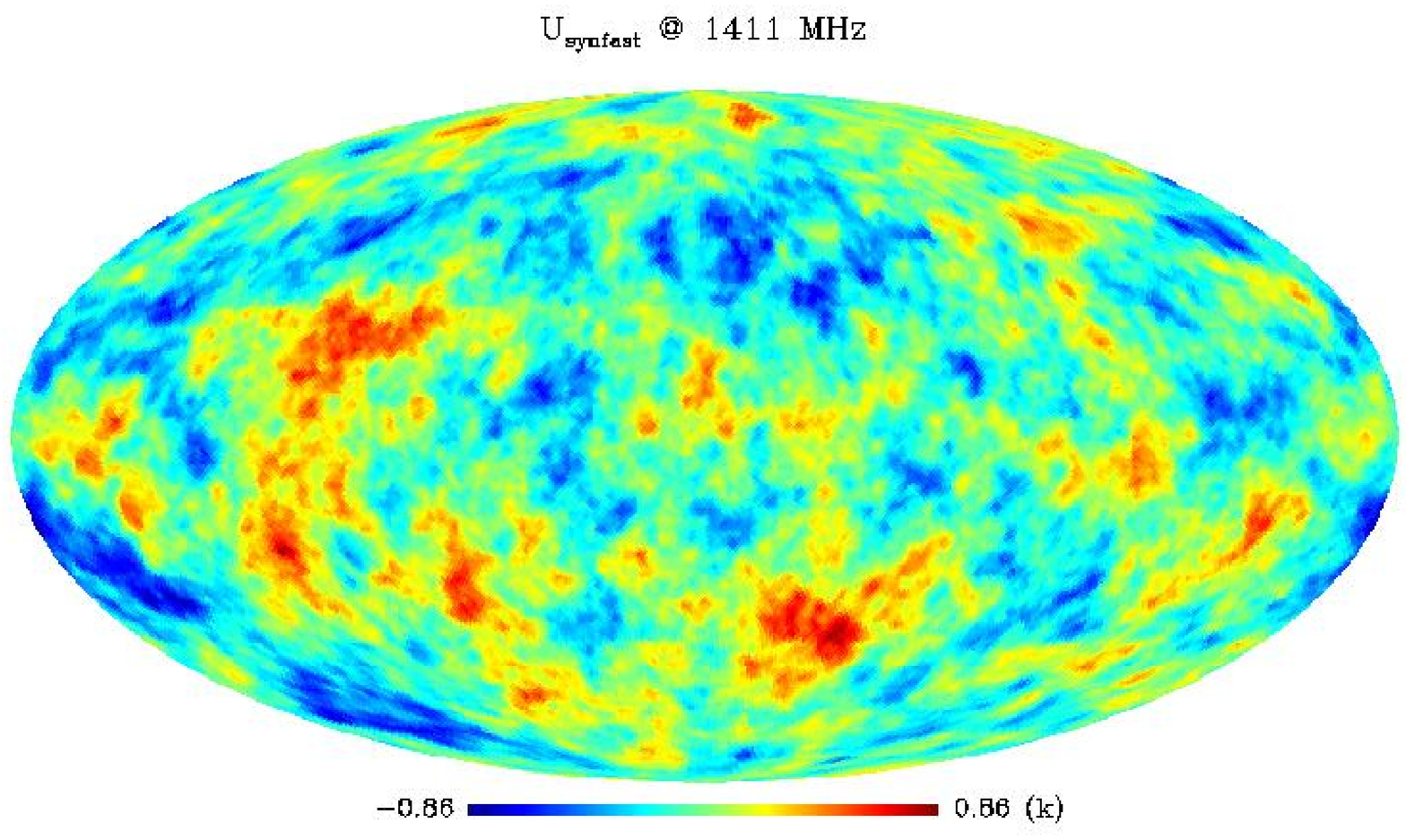}&
   \includegraphics[width=5cm,angle=90,clip=]{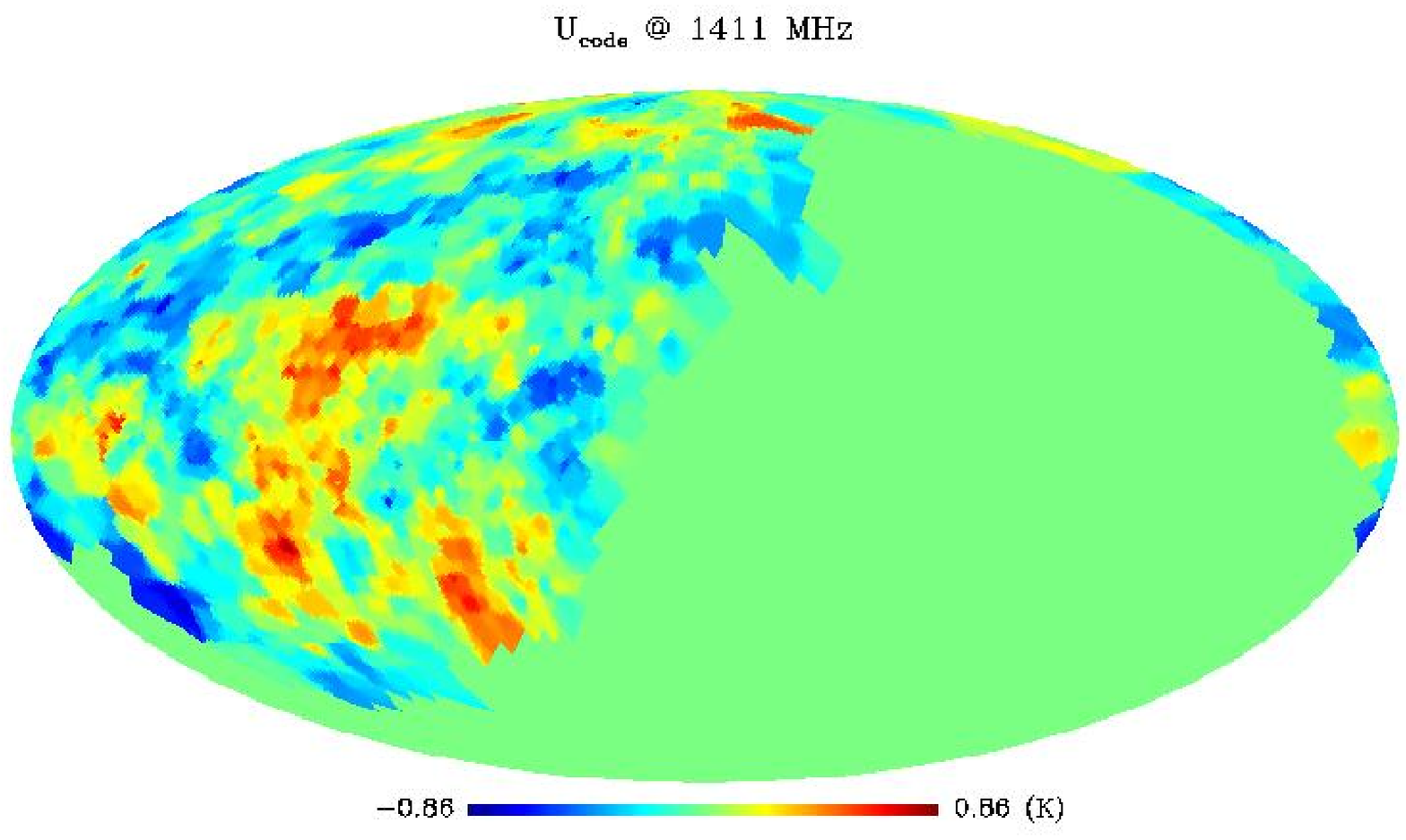}\\
   \end{tabular}
   \caption{{\sc Synfast} realizations (left side) at 1411~MHz
versus maps reconstructed applying the final version of
the filling code to the synthetic data table (right side). See 
text for details.}
   \label{synfvscodemaps}
 \end{figure*}

In order to evaluate and optimize the signal reconstruction quality
provided by our algorithm, we performed a final 
test. Starting from simulated
polarization maps having known statistical properties,
we created a table of data analogous to the Leiden
survey ones and run our code to build
$PI$, $Q$ and $U$ maps to be compared with the input skies.
We focused on the 1411~MHz case, because this
is the most seriously undersampled both in terms of
number of observations per beam and of observed
points.
Therefore, the maps produced at this frequency
are potentially most influenced by the
interpolation method effects, particularly
at multipoles larger than some tens.
We assumed
$C_{\ell}^{E}=C_{\ell}^{B}=C_{\ell}={\kappa}\cdot\ell~^{\alpha}$,
with values of the parameters ${\kappa}$ and ${\alpha}$
in agreement with the mean law obtained for the Galactic polarized synchrotron
emission in our first analysis of the 1411~MHz survey
(reported in \cite{buriganalaporta02}). 
 For $\ell \sim [30,200]$ we adopted~\footnote{
For lower multipoles we assumed a slope $\le -2.5$. Precisely, in this 
test we used $\alpha = -2.5$ for $\ell \sim [6,30]$ 
and $\alpha = 0$ for $\ell < 6$.} 
$\kappa = 1 (K^2)$ and $\alpha = -3$.
Applying the HEALPix facility {\sc Synfast} to these angular power
spectra,
we generated maps
of $Q$ and $U$ with a pixel size of $\simeq 13.7'$
and a beamwidth $\theta_{HPBW}=36'$, as in the 1411~MHz survey,
and used them to build the simulated table of data as described
below.\\
At each observed position of the Leiden survey
\begin{itemize}
\item[a)] We first
assigned the Stokes parameters corresponding to the {\sc Synfast}
         simulation pixel in which it would project and
\item[b)] The corresponding rms given in the original table.
\item[c)] We then added to each
Stokes parameter a random value extracted from
a Gaussian distribution
in order to mimic the effect of the instrumental white
noise (see, e.g., \cite{buriganasaez03}).
The adopted standard deviation of this Gaussian distribution has been
chosen equal to the mean error quoted for the considered  original data.
In this way the signal-to-noise ratio of the simulated table
is statistically very close to the real case one.
\end{itemize}
We then obtained $PI$ as $\sqrt{Q^2+U^2}$ and calculated
its error $\sigma_{PI}$ according to the standard error propagation rules.\\
Finally, we applied our interpolation method to the simulated data table
and produced independently $PI$ ,$Q$, and $U$ simulated maps with
$\theta_{pix} \simeq 0.92^{\circ}$.
We ran our code several times, changing the weight $n$ to be adopted
in the average and the criteria for selecting the interpolation radius,
so to sort out the best solution.\\
As a first self-consistency test of the resulting maps,
we verified that typically the $PI$ map can be considered equal to the
$\sqrt{Q^2+U^2}$  within the uncertainty limit in $85\%$ of the
pixels.
Then we compared the reconstructed maps with the input sky maps
directly obtained
with the {\sc Synfast} facility (see Fig.~\ref{synfvscodemaps}).

\begin{figure*}
\centering
\includegraphics[width=6cm,angle=90]{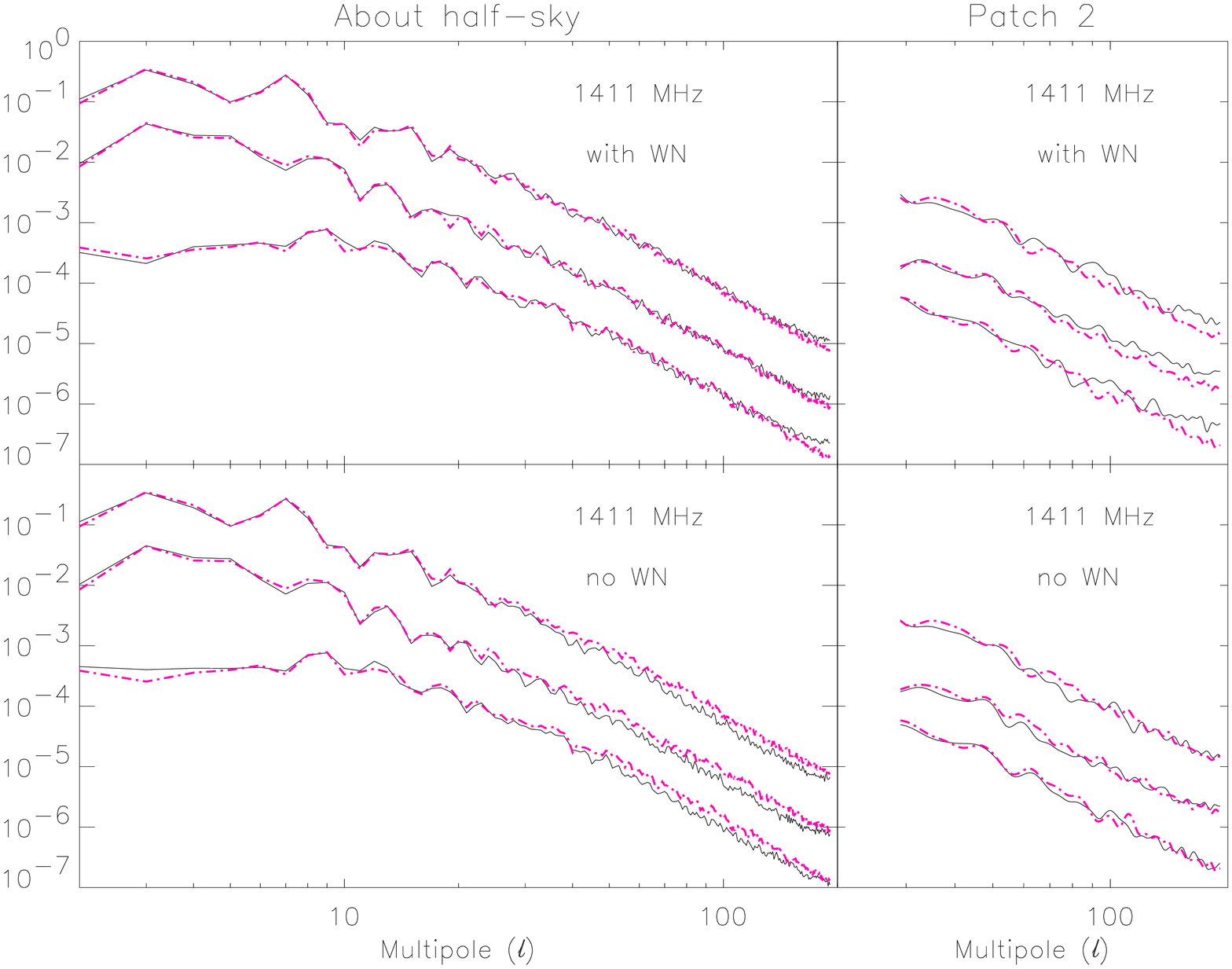}
\vskip 0.5cm
\caption{Comparison between the APSs 
of the input maps 
(dot lines) extracted
from the {\sc Synfast} simulations (degraded to 
$0.92^{\circ}$ pixel size maps) 
and the reprocessed maps (solid lines). 
From the bottom: $C_{\ell}^{PI}$, 
$C_{\ell}^{E}\cdot10$ and  $C_{\ell}^{B}\cdot100$.  
Note the excellent agreement between the input and reprocessed
APSs up to $\sim 100$.}
\label{apscomparison}
\end{figure*}

We first degraded~\footnote{We verified that this
operation does not significantly affect the APS on the multipole
range of interest.} the {\sc Synfast} simulations
to $0.92^{\circ}$ pixel size maps
and considered only the portion of the sky
covered by the Leiden survey.
Computing the APSs of both the original and the reprocessed maps
we found that the best agreement between them
is obtained:

\begin{enumerate}
\item Adopting as polarized intensity the quantity 
 $\sqrt{Q_{reproc}^2+U_{reproc}^2}$ 
(instead of the $PI_{reproc}$, produced applying the filling code 
to the $PI$ values of the data table); 
\item Using the weight $1/(\sigma^{2}\cdot d^{4})$; and
\item Interpolating the data in the minimum circle containing at least two 
observations.   
\end{enumerate}
The first choice is motivated by the necessity to 
(implicitly) preserve the information on the 
polarization angle and is a posteriori 
justified by the following result: the APS
of $PI$ map reconstructed 
applying the code directly on the table of $PI$ values 
presents a certain loss in power with respect to the input case.
The reason is likely that the $PI$ interpolated value  
overestimates the effective mean value of the field 
in the presence of changes in the signs of the Stokes 
parameters and consequently leads to an underestimation of the
 mean value of the field fluctuations.\\
Criteria 2 and 3 imply that the interpolation is as 
local as possible according to the sampling of each sky region. 
This approach is quite different from that pursued by other authors 
(\cite{bruscoli02}); they performed a Gaussian 
smoothing everywhere with constant $\theta_{HPBW}$ (of $3^\circ$), neglecting 
the sampling dependence on the position in the sky and 
on the frequency.\\  
In principle, before computing any average we should have transformed 
the polarization vectors using the formula of the parallel transport on 
the sphere;
however operating only a local interpolation, the consequent
mixing of the two orthogonal components of the polarized signal turned out to 
be negligible~\footnote{In order to drag vectors on the 
sphere preserving strength
and angles, one should apply the parallel transport formula
estabilished by differential geometry.
Integrating on paths close to geodetic, a suitable approximation
is given by \\
$p_{\theta}^{(PT)} = p_{\theta} \cdot \cos(t) + p_{\phi} \cdot \sin(t)$\\
$p_{\phi}^{(PT)} = - p_{\theta} \cdot \sin(t) + p_{\phi} \cdot \cos(t)$\\
where $t = cos(\theta_{ave}) \cdot \delta\phi$ and
 $\theta_{ave}=(\theta_{in}+\theta_{fin})/2$
(see \cite{bruscoli02} for further details).
Using these equations it is possible to estimate the percentage
contribution of each initial Stokes parameter to the new parameter expression,
once it has been moved to another point of the sphere. We
assumed $\delta\phi=3^{\circ}$, corresponding to the mean value of the
interpolation radius in the selected patches. Neglecting
the parallel transport in our interpolation method causes a mixing of 
the Stokes parameters of $\simeq 1\%$ in the equatorial area and of  
$\simeq 5\%$ in the medium and high
latitude regions.}.\\
Fig.~\ref{apscomparison} shows that the statistical information contained in the 
synthetic data table is appreciably traced out in the
multipole ranges of interest both for the half-sky map and for the 
patches (for conciseness we report here 
only the results found for patch 2).

\begin{figure*}
\centering
\includegraphics[width=7cm,angle=90]{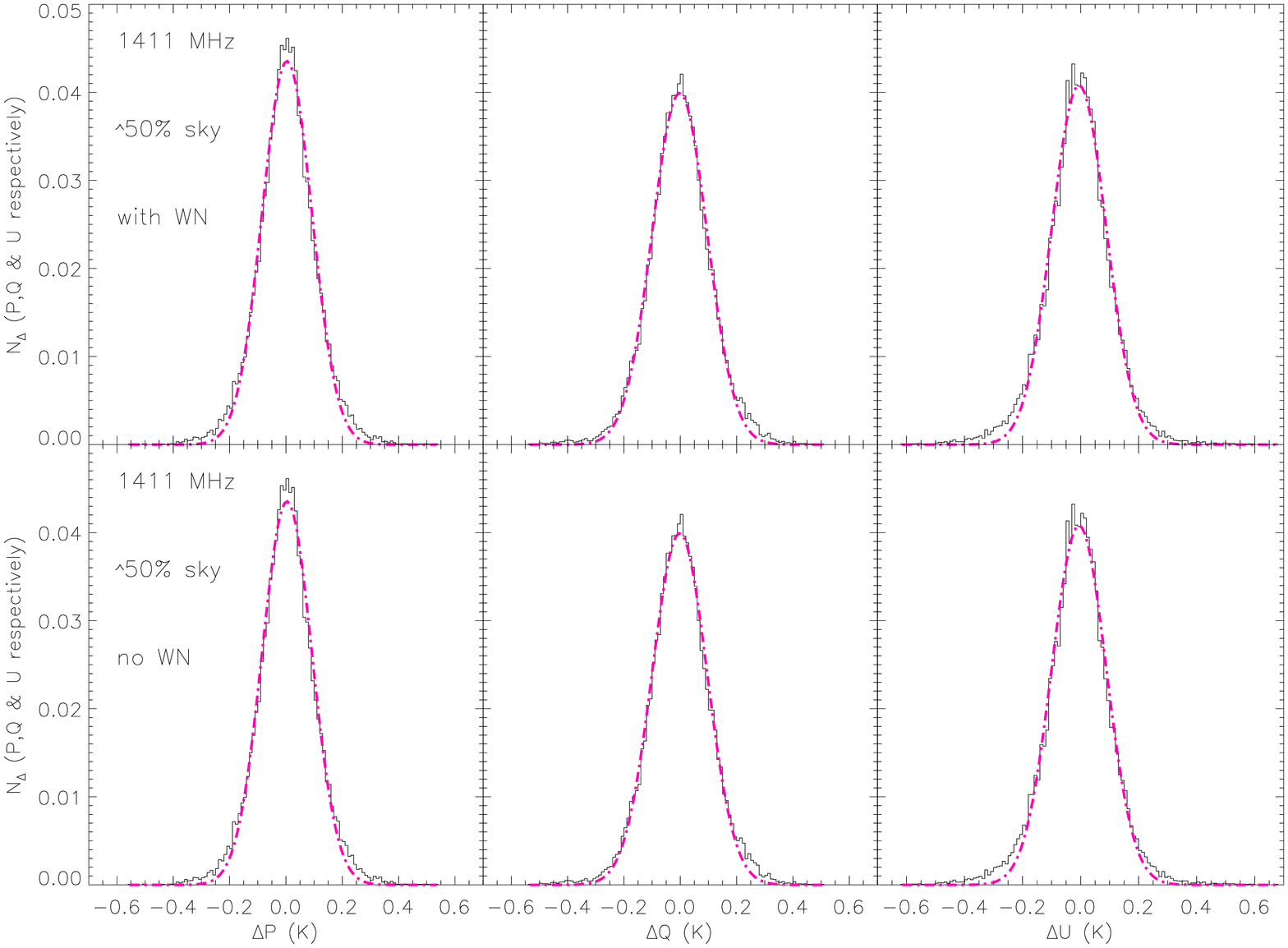}
\vskip 0.5cm
\caption{Distribution of the pixel to pixel differences between the 
maps directly simulated with {\sc Synfast} 
(degraded to $0.92^{\circ}$ pixel size maps)
and the reprocessed maps at 1411~MHz. We display here the results of the
test, both with (upper panels) and without (lower panels) white noise 
contamination, in the case of the full coverage 1411~MHz Leiden survey. 
The superimposed curves are the Gaussian distributions giving the 
best fit to the data. } 
\label{GaussPlots_1411}
\end{figure*}

\noindent
Fitting the APS of the reprocessed map, we
quantified the precision level at which the input values of
the amplitude and of the spectral index were recovered:
by considering several realizations for the various cases,
we found relative errors within $\simeq 10\%$ for both the parameters.
\\
We also checked the reliability of the maps themselves
in a more quantitative way. 
The distributions of the pixel to pixel difference 
between the two series of maps 
(i.e. $PI$, $Q$ and $U$ produced by {\sc Synfast} and reprocessed
by our interpolation code) are, 
in all cases, close to Gaussian distributions, 
with standard deviations comparable with the typical mean 
error of the data table (see Table \ref{DiffDistr_1411} and 
Fig. \ref{GaussPlots_1411}). 

\begin{table*}
\begin{center}
\begin{tabular}{|c|c|c|c|c|c|c|c|}
\hline 
{Half-sky} & \multicolumn{2}{c|}{Gaussian interp.} & 
            \multicolumn{2}{c|}{distribution} &
           \multicolumn{3}{c|}{\% of pixels} \\
&  {ave (mK)} & {rms (mk)} & {ave (mK)} & {rms (mk)} &
$1\sigma$  & $2\sigma$  & $3\sigma$  \\
\hline
$\Delta PI$ & 3.6 & 88.3 & 3.9 & 102.2 &  71.8 & 94.7 & 99.0 \\
\hline
$\Delta Q$ & -2.4 & 97.1 & -0.9 & 109.6 & 71.5 & 94.9 &99.1\\
\hline
$\Delta U$ & -5.6 & 93.5 & -9.1 & 114.3 & 73.7 & 94.8 & 98.7 \\
\hline
\hline
{Patch 1} & \multicolumn{2}{c|}{Gaussian interp.} & 
            \multicolumn{2}{c|}{distribution} &
           \multicolumn{3}{c|}{\% of pixels} \\
&  {ave (mK)} & {rms (mk)} & {ave (mK)} & {rms (mk)} &
$1\sigma$  & $2\sigma$  & $3\sigma$  \\
\hline
$\Delta$P & 6.3 & 61.6  & 7.7 & 73.7 &  71.8 & 94.4 & 98.9 \\
\hline
$\Delta$Q & -7.5 & 68.2 & -8.2  & 76.2 & 70.8 & 95.4 & 99.2 \\
\hline
$\Delta$U & -6.6 & 66.1 & -6.8 & 77.4 & 72.1 & 94.4 & 99.1 \\
\hline
\hline
{Patch 2} & \multicolumn{2}{c|}{Gaussian interp.} & 
            \multicolumn{2}{c|}{distribution} &
           \multicolumn{3}{c|}{\% of pixels} \\
&  {ave (mK)} & {rms (mk)} & {ave (mK)} & {rms (mk)} &
$1\sigma$  & $2\sigma$  & $3\sigma$  \\
\hline
$\Delta PI$ & 1.2 & 74.9 & 3.5 & 80.3 & 70.7 & 95.1 & 99.6 \\
\hline
$\Delta Q$ & 1.6 & 81.1 & -3.5 & 85.5 & 70.0 & 94.8 & 99.4 \\
\hline
$\Delta U$ & -3.2 & 75.0 & -10.2 & 85.2 & 70.5 & 95.9 & 99.0 \\
\hline
\hline
{Patch 3} & \multicolumn{2}{c|}{Gaussian interp.} & 
            \multicolumn{2}{c|}{distribution} &
           \multicolumn{3}{c|}{\% of pixels} \\
&  {ave (mK)} & {rms (mk)} & {ave (mK)} & {rms (mk)} &
$1\sigma$  & $2\sigma$  & $3\sigma$  \\
\hline
$\Delta PI$ & 6.0 & 72.8 & 1.4 & 78.9 & 69.4 & 96.3 & 99.7 \\
\hline
$\Delta Q$ & 1.2 & 68.5 & -3.4 & 71.9 & 69.0 & 94.6 & 99.7 \\
\hline
$\Delta U$ & -13.9 & 75.7 & -13.9 & 81.0 & 68.0 & 94.6 & 99.3 \\
\hline
\hline
\end{tabular}
\end{center}
\caption{The pixel to pixel difference between the 
maps directly generated with {\sc Synfast} (after degradation to
$0.92^\circ$ pixel size maps) and the reconstructed maps contaminated by 
white noise,
for the whole survey and the considered patches.
Columns 2 and 3 give the average and rms of the Gaussian 
function that best fits the difference distribution (shown
	  in Fig.~\ref{GaussPlots_1411}).
Columns 4 and 5 contain the average and standard deviation
of the difference distribution. 
Columns 6, 7 and 8 give the percentages of pixels with 
absolute difference within the reported levels (here $1\sigma$ 
corresponds to the rms reported in column 5). } 
\label{DiffDistr_1411}
\end{table*}
%

\noindent
We repeated the test described above 
skipping the 
step c)~\footnote{
Formally, in this case we apply the step c) but using
in the noise generation a standard deviation 
 negligible with respect to the values of $PI$, 
$Q$ and $U$ (the implemented code assumes that a mean error is 
given for each measurement of the observed quantities).}.
The distributions of the differences between the input and reprocessed 
maps are again close to Gaussian distributions, but with 
standard deviations slightly smaller than those found including step c)
(in particular, for the full survey coverage 
the standard deviation is now smaller by about $20\%$).
This indicates the existence of a {\it geometric} contribution in the
error distribution over the maps related to the poor sampling 
of the data. Expressing the rms of the pixel to pixel 
difference distribution  as
$$ \sigma_{map} \simeq \sqrt{\sigma_{samp}^2+\sigma_{WN}^2} $$ 
the analysis of reprocessed maps 
(with and without noise) gives
$$ \sigma_{WN} \simeq {\rm const} \times \sigma_{samp} $$
with ${\rm const} \simeq 0.63, 0.87, 0.90, 1.15$, respectively
for the whole survey coverage and for the patches 1, 2, and 3,
without significant differences among the maps
of $PI$, $Q$ or $U$.\\
The relative weight of the geometrical contribution to the 
overall error decreases with the goodness of the sampling.
At the other frequencies of the Leiden surveys the sampling and the 
signal-to-noise ratio are better than 
at 1411~MHz; as a 
consequence, we can look at the 
ratio between  $\sigma_{samp}$ (resp. $\sigma_{map}$)
 and the signal at 1411~MHz as a conservative upper 
limit to the relative 
error caused by geometric effects 
(resp. geometric effects plus experiment white noise) 
at the lower frequencies.

To verify whether the goodness of the signal reconstruction depends on 
the input sky characteristics, we repeated the entire test adopting
a different input APS, more appropriate at 408~MHz.
At this frequency the analysis of the preliminary polarization maps led
to different results for the APSs than the 1411~MHz case.
In particular, the spectral index was flatter (roughly $\sim -1.5$
in the multipole range [30,200]), 
both for the whole coverage map and for the patches.
The same test repeated with this slope confirmed the results
described before.
The quality of the APS reconstruction is 
slightly less accurate than in the previous case, however the 
relative errors estimated for the recovered amplitudes and 
spectral indices remain within $\simeq 10 \%$. \\

\section{Results}

\begin{figure*} 
   \centering
   \begin{tabular}{ccc}
   \includegraphics[width=3cm,angle=90,clip=]{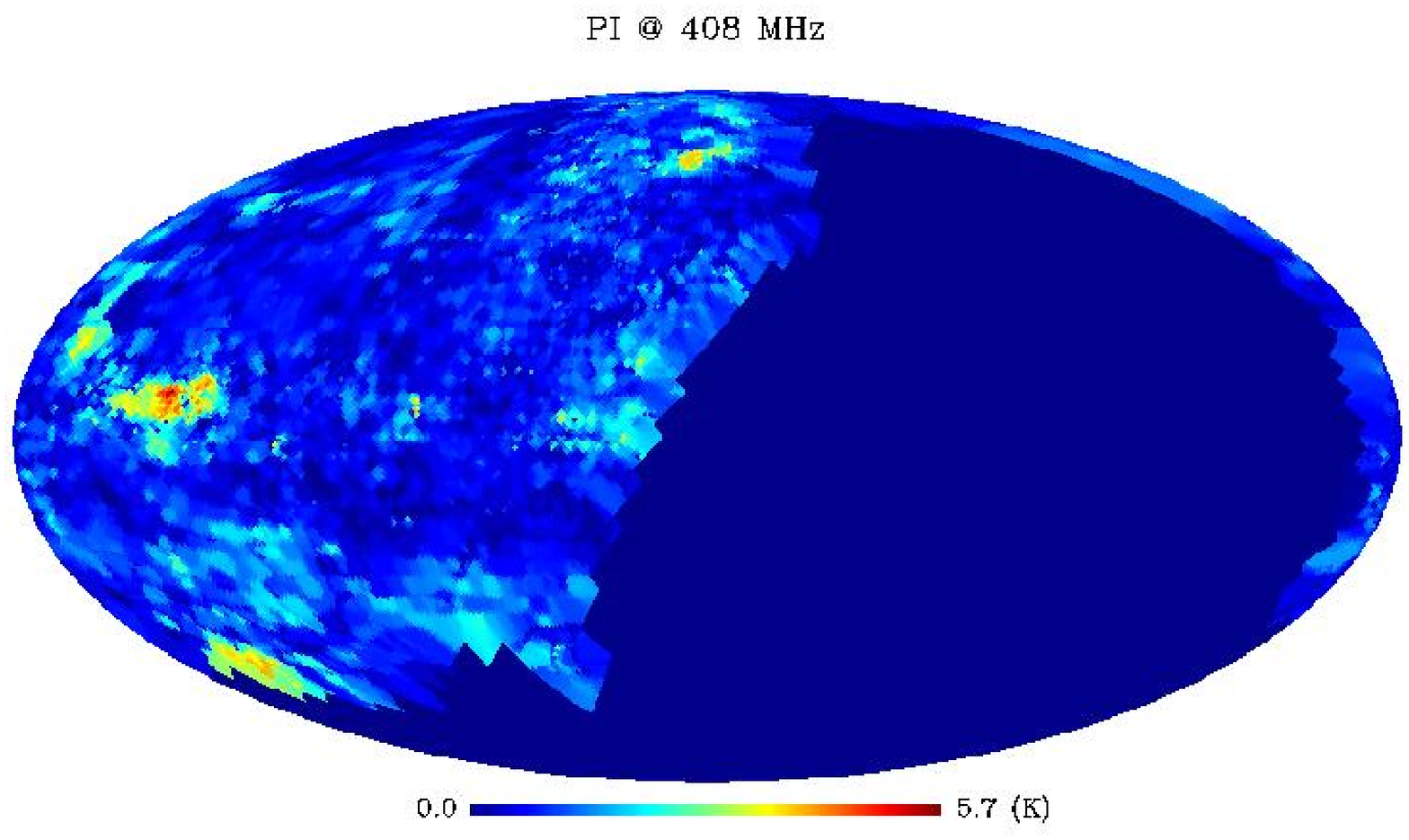}&
   \includegraphics[width=3cm,angle=90,clip=]{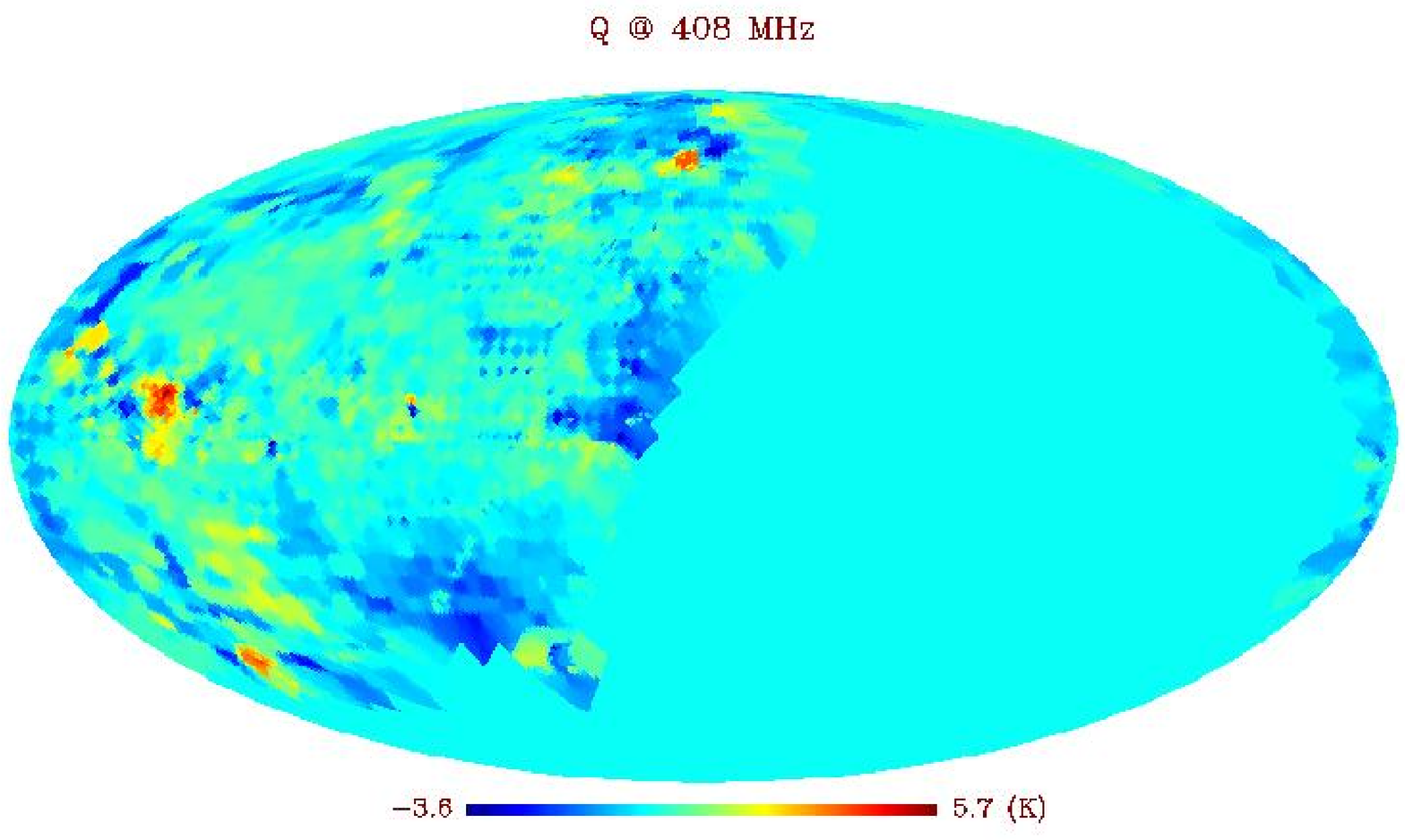}&
   \includegraphics[width=3cm,angle=90,clip=]{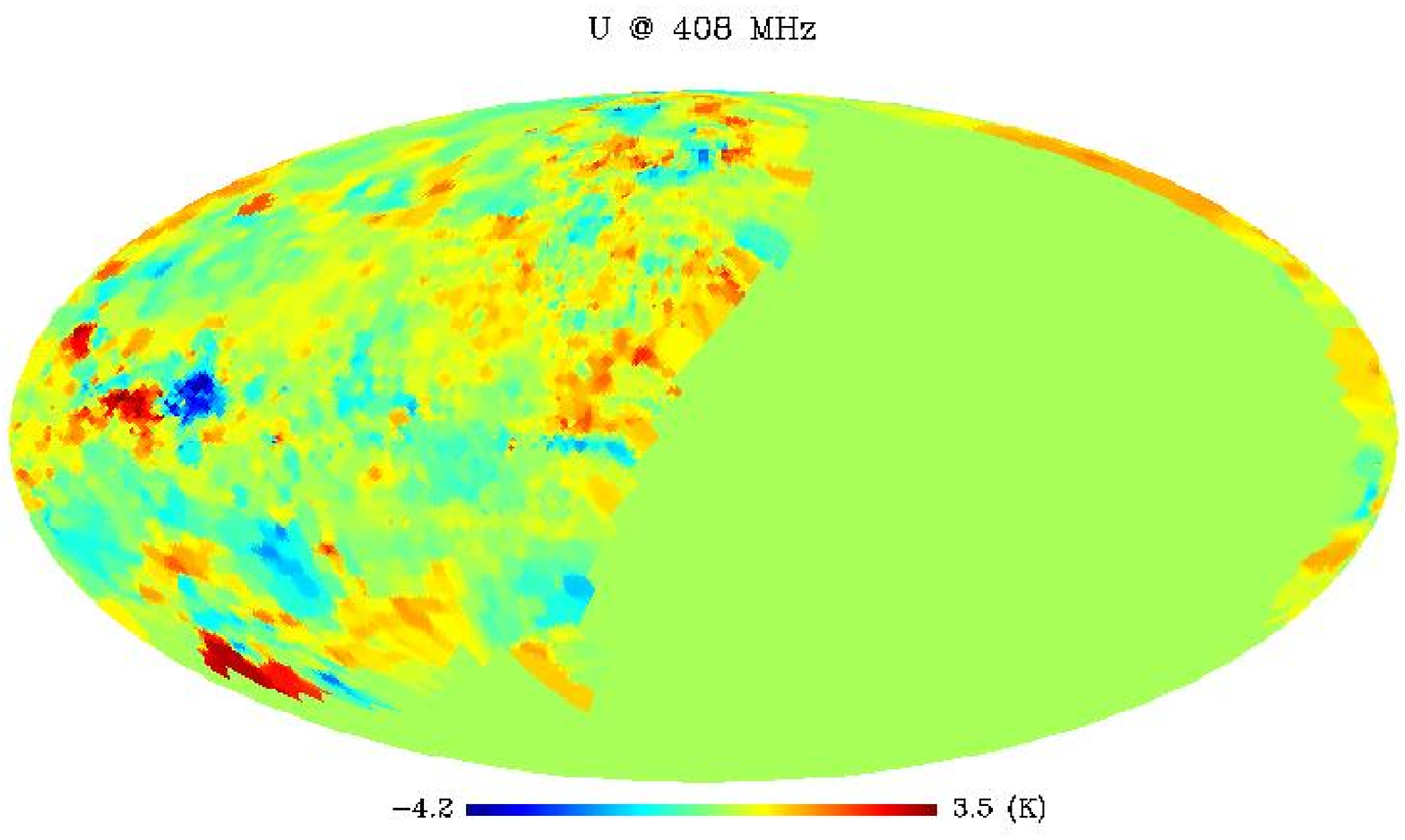}\\
   \includegraphics[width=3cm,angle=90,clip=]{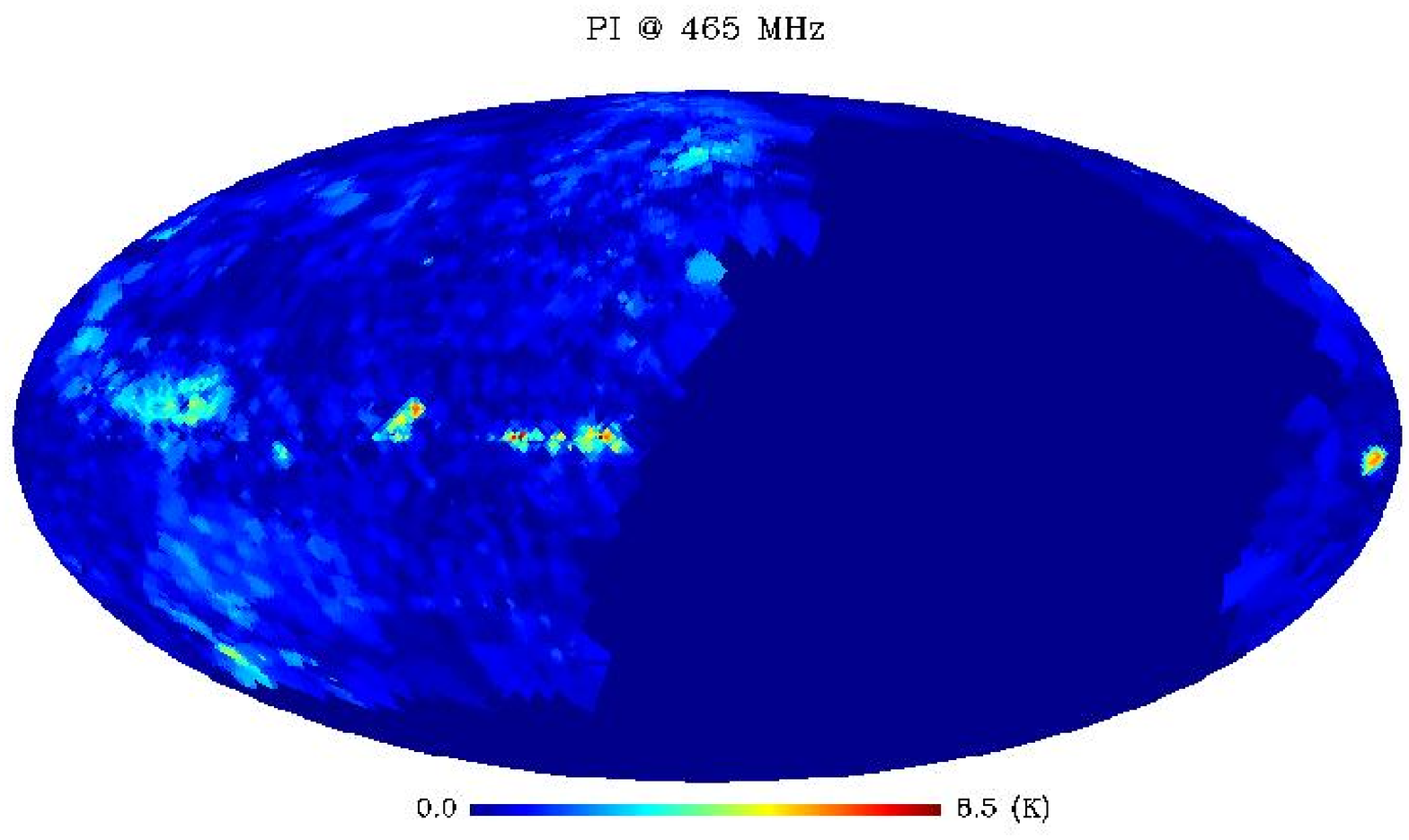}&
   \includegraphics[width=3cm,angle=90,clip=]{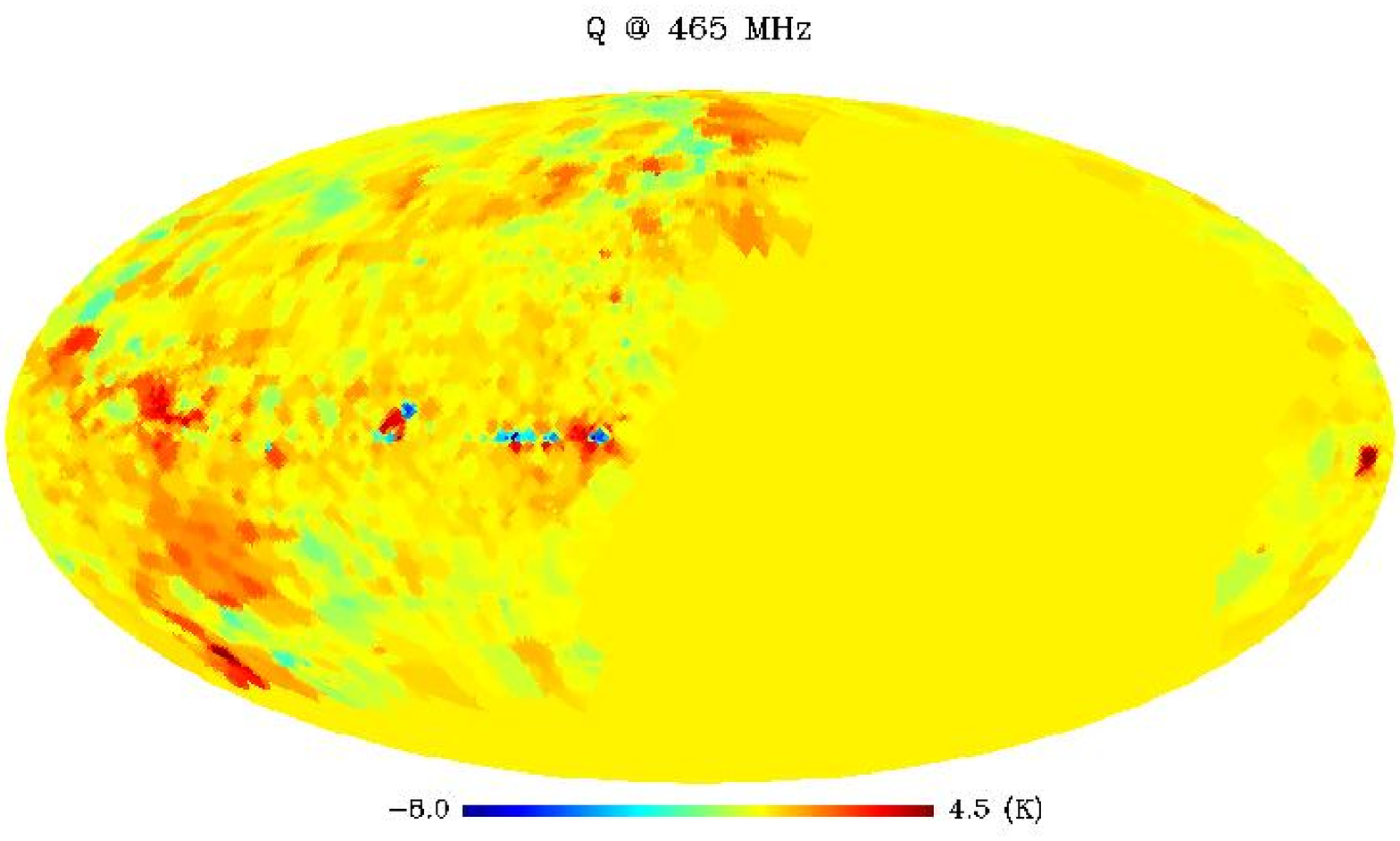}&
   \includegraphics[width=3cm,angle=90,clip=]{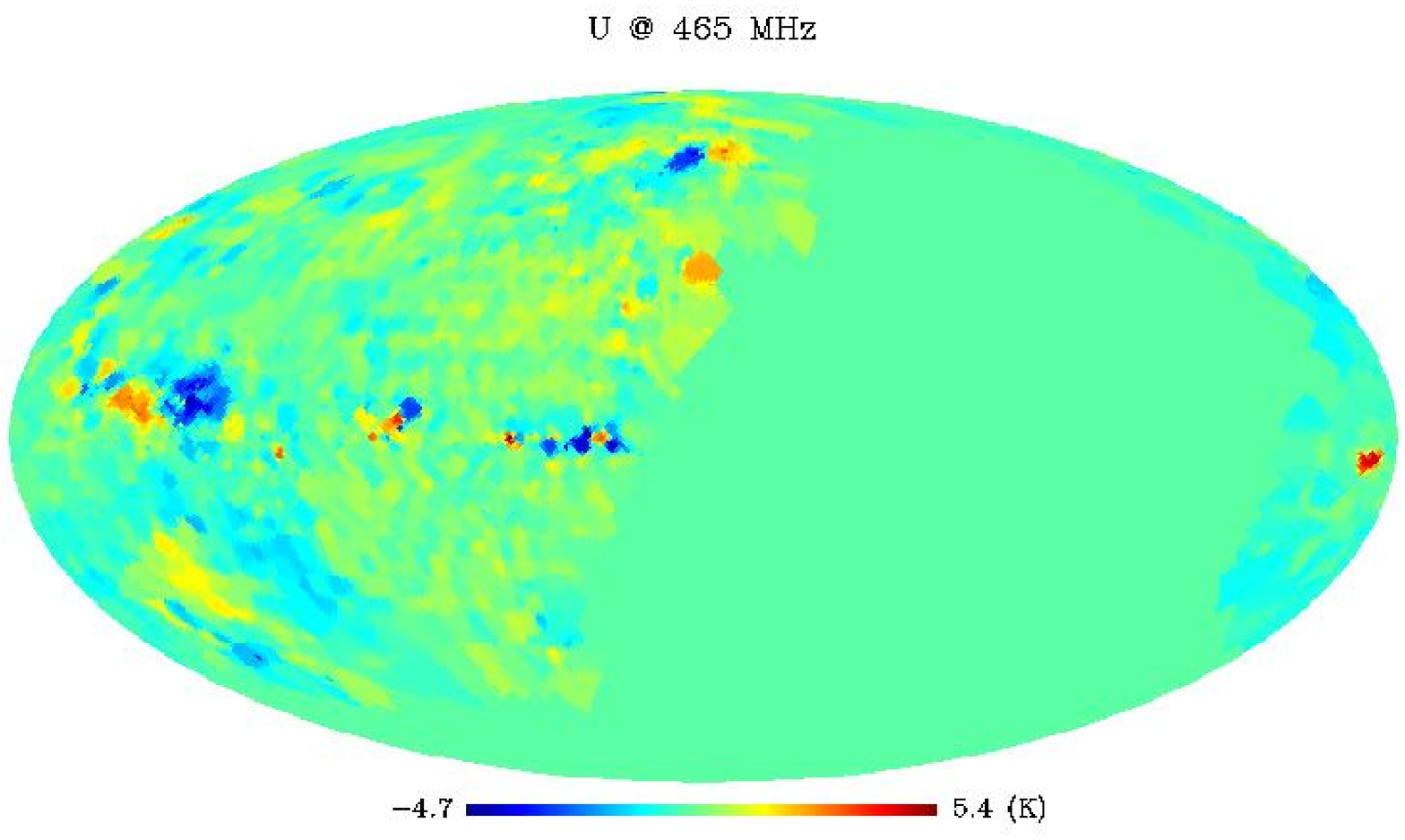}\\
   \includegraphics[width=3cm,angle=90,clip=]{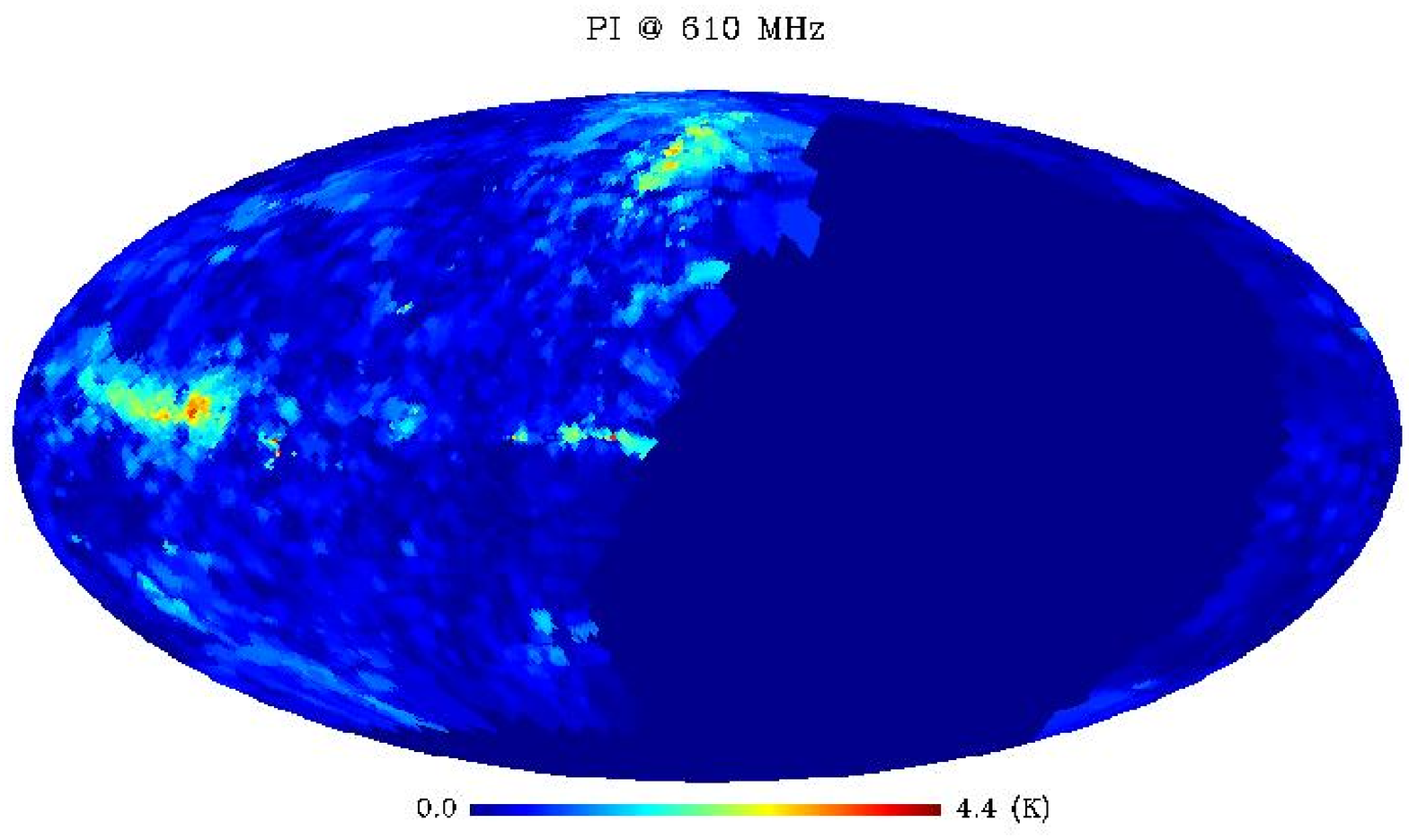}&
   \includegraphics[width=3cm,angle=90,clip=]{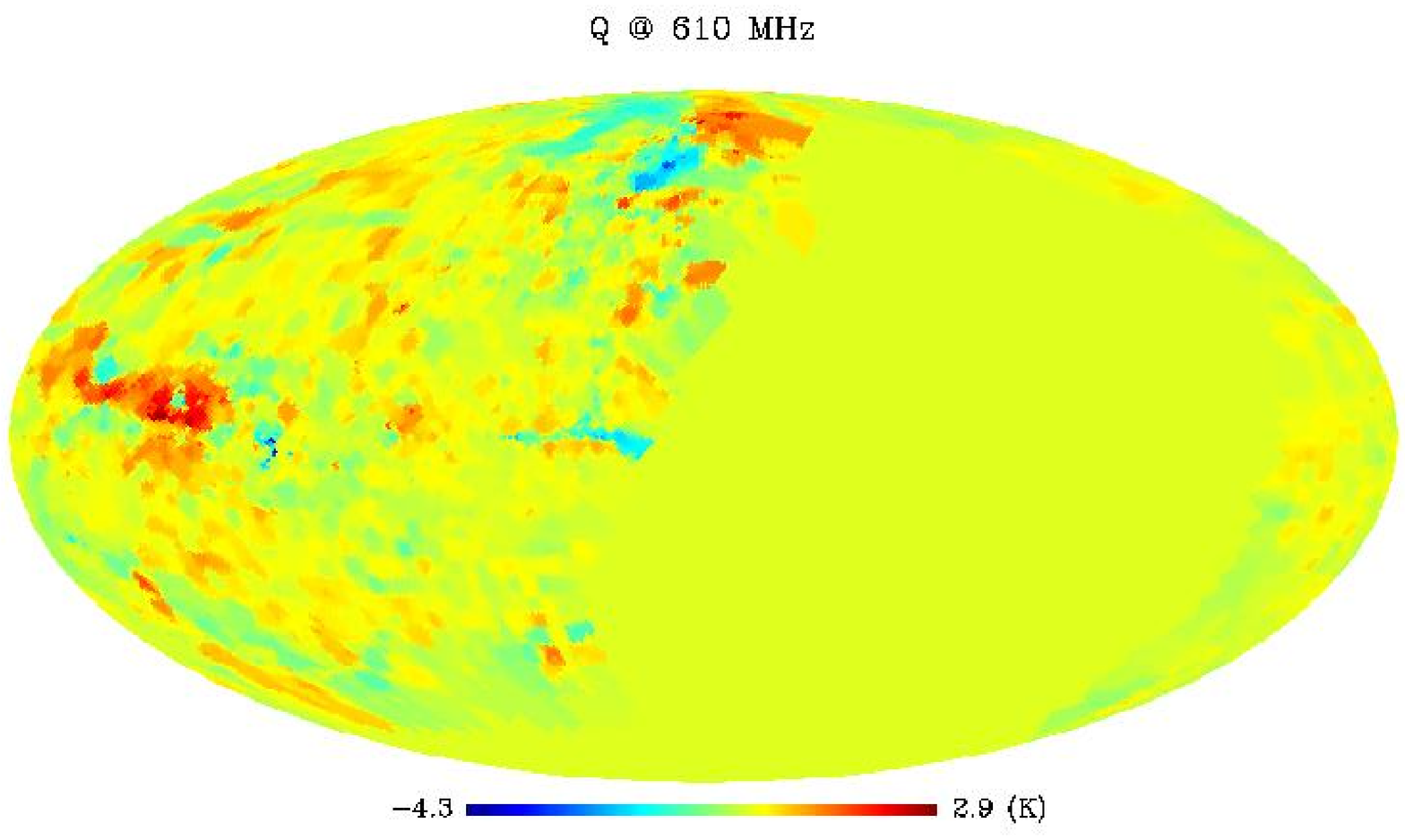}&
   \includegraphics[width=3cm,angle=90,clip=]{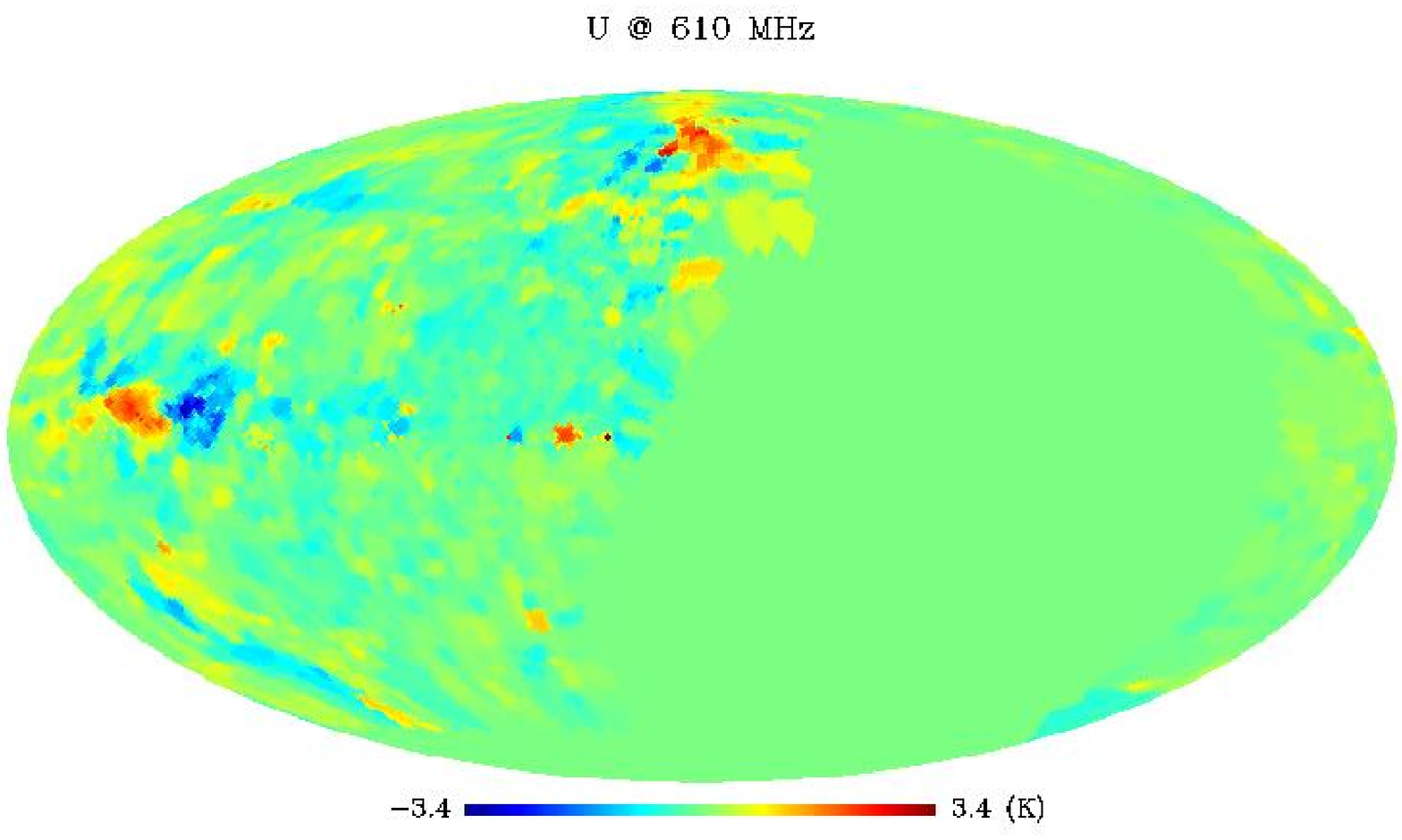}\\
   \includegraphics[width=3cm,angle=90,clip=]{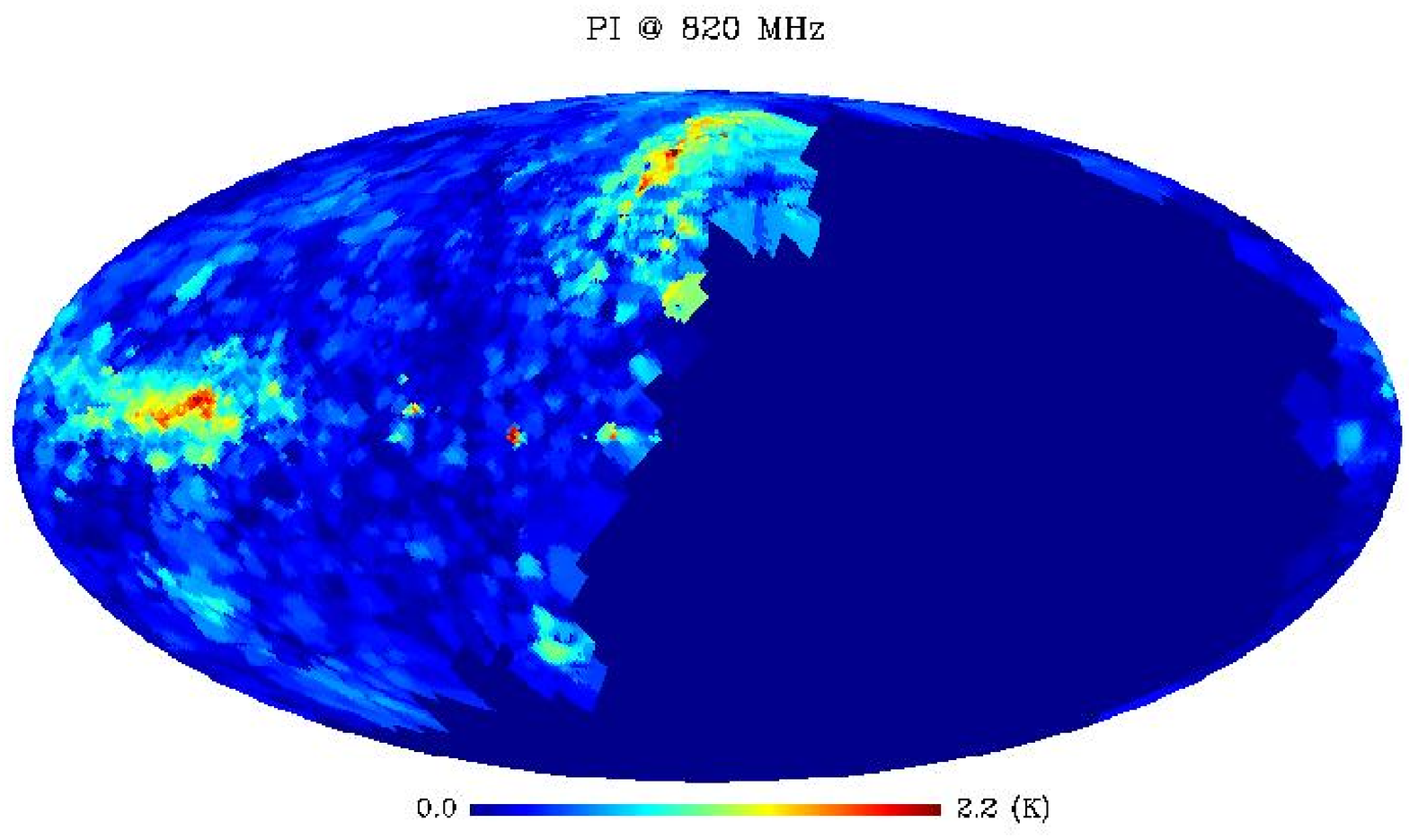}&
   \includegraphics[width=3cm,angle=90,clip=]{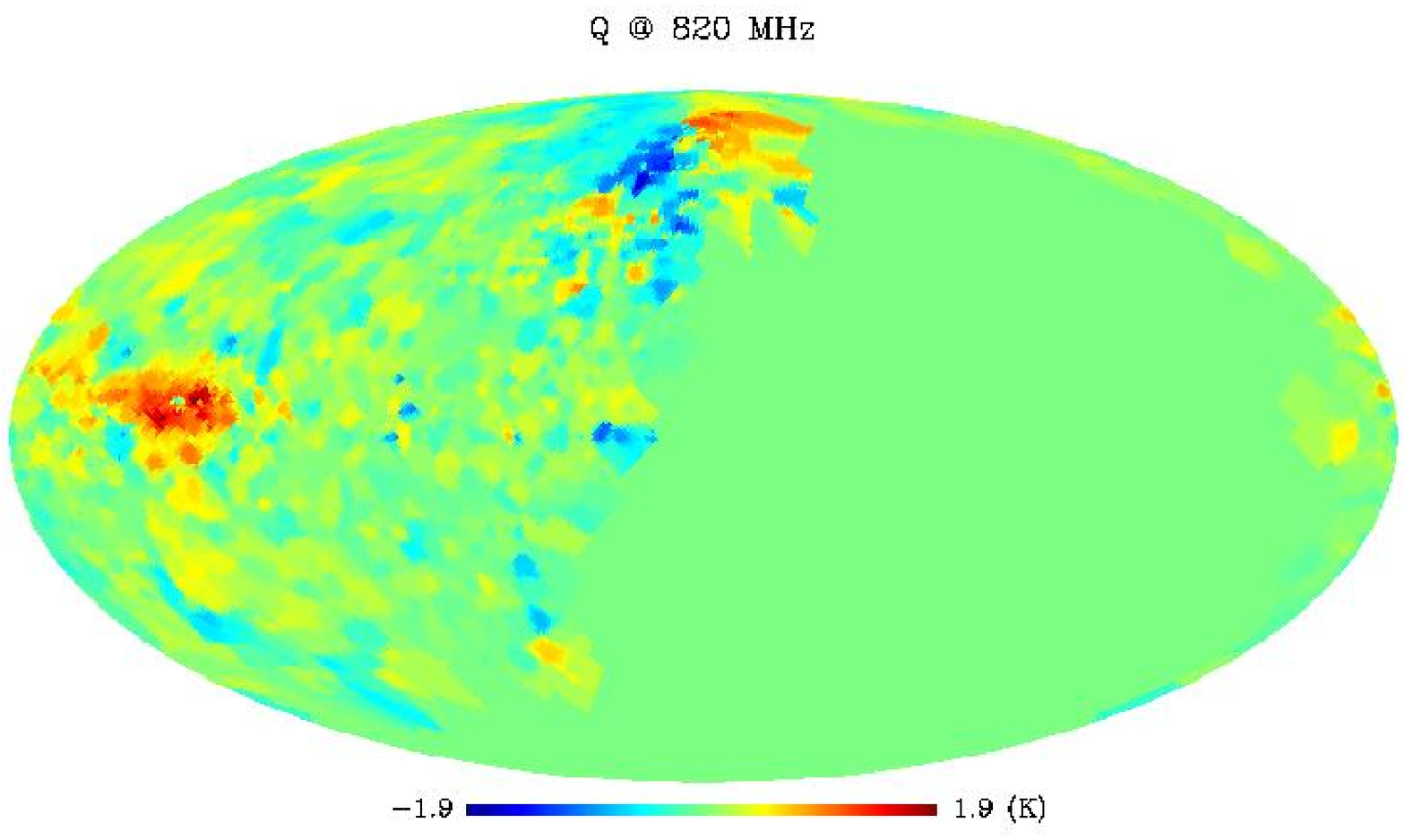}&
   \includegraphics[width=3cm,angle=90,clip=]{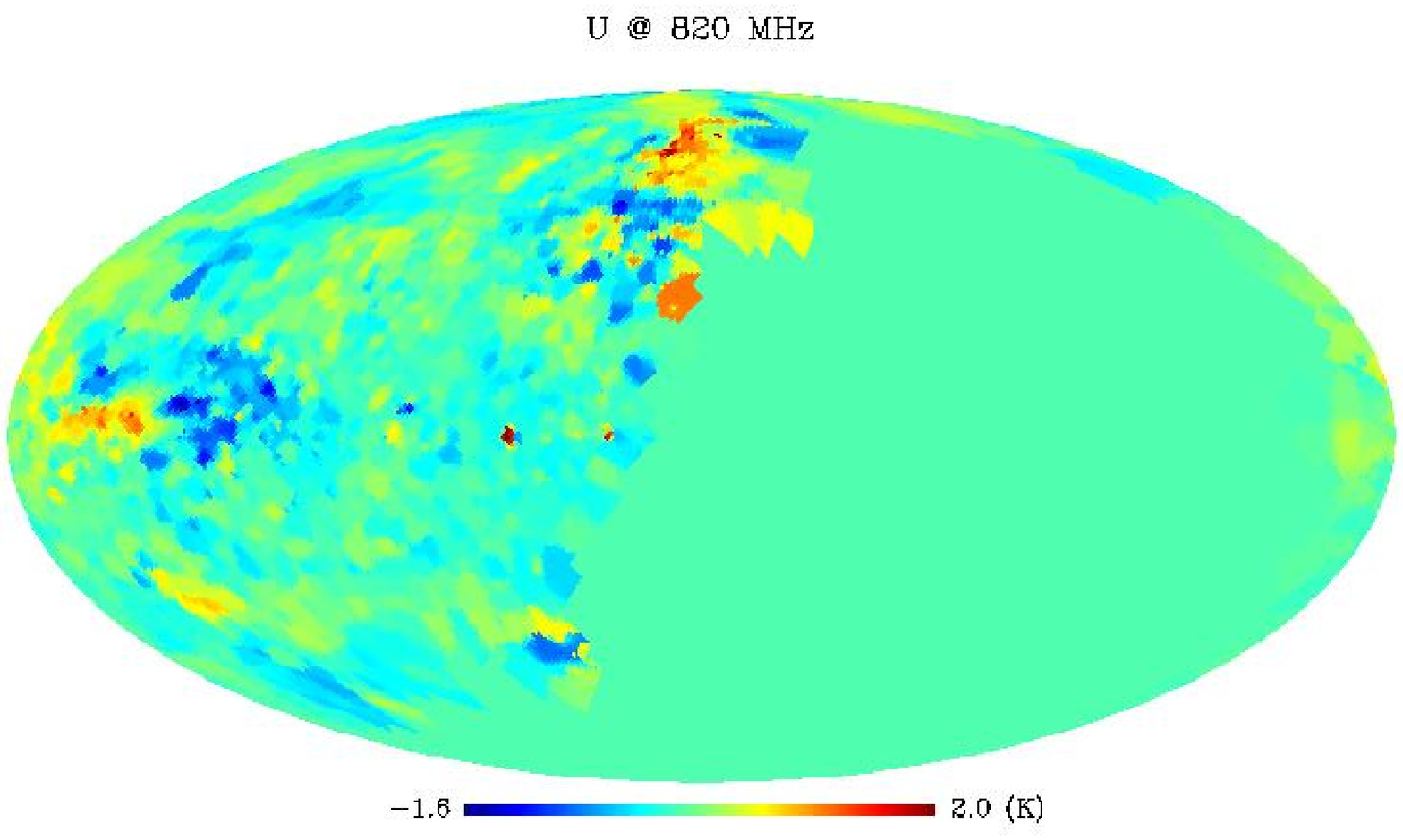}\\
   \includegraphics[width=3cm,angle=90,clip=]{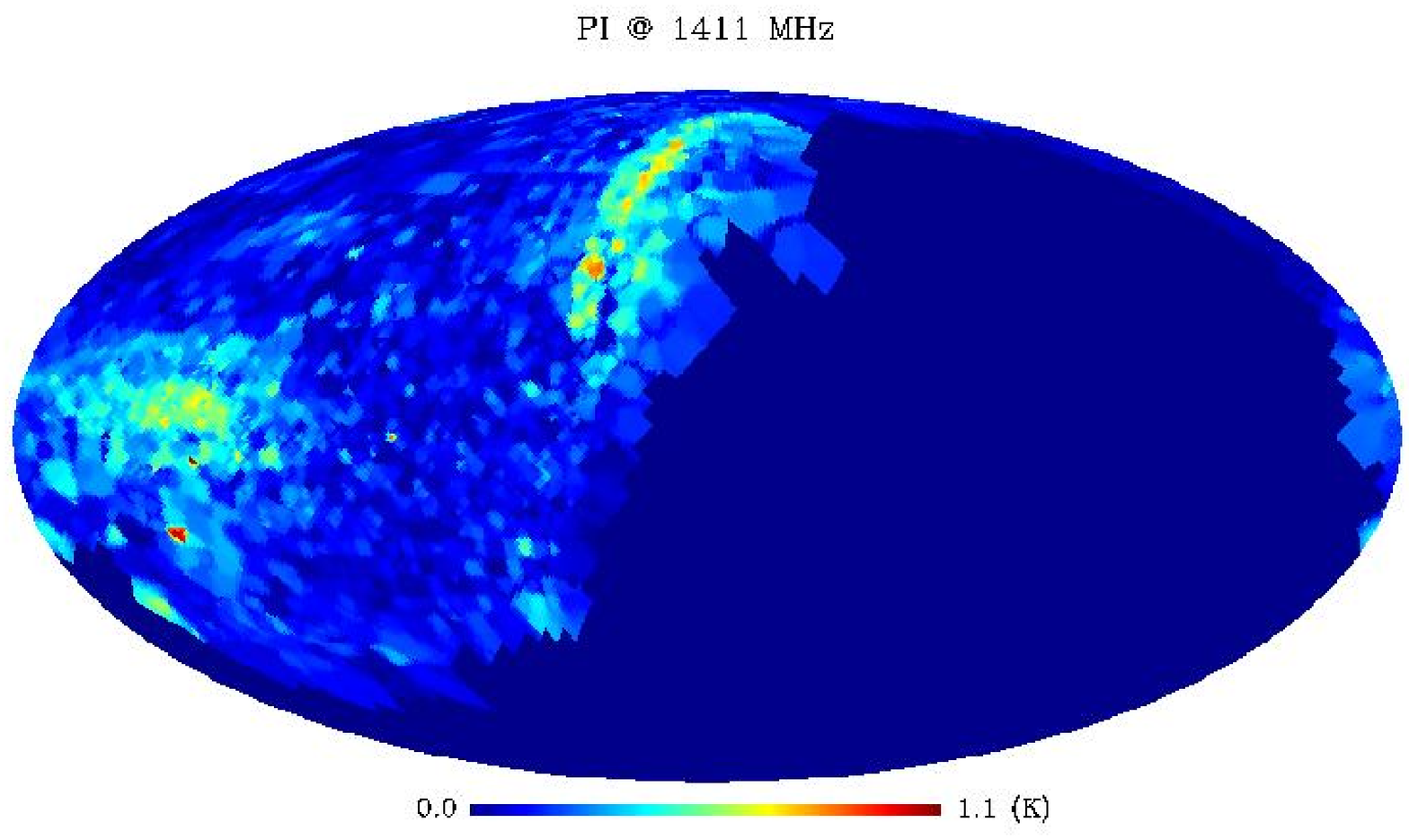}&
   \includegraphics[width=3cm,angle=90,clip=]{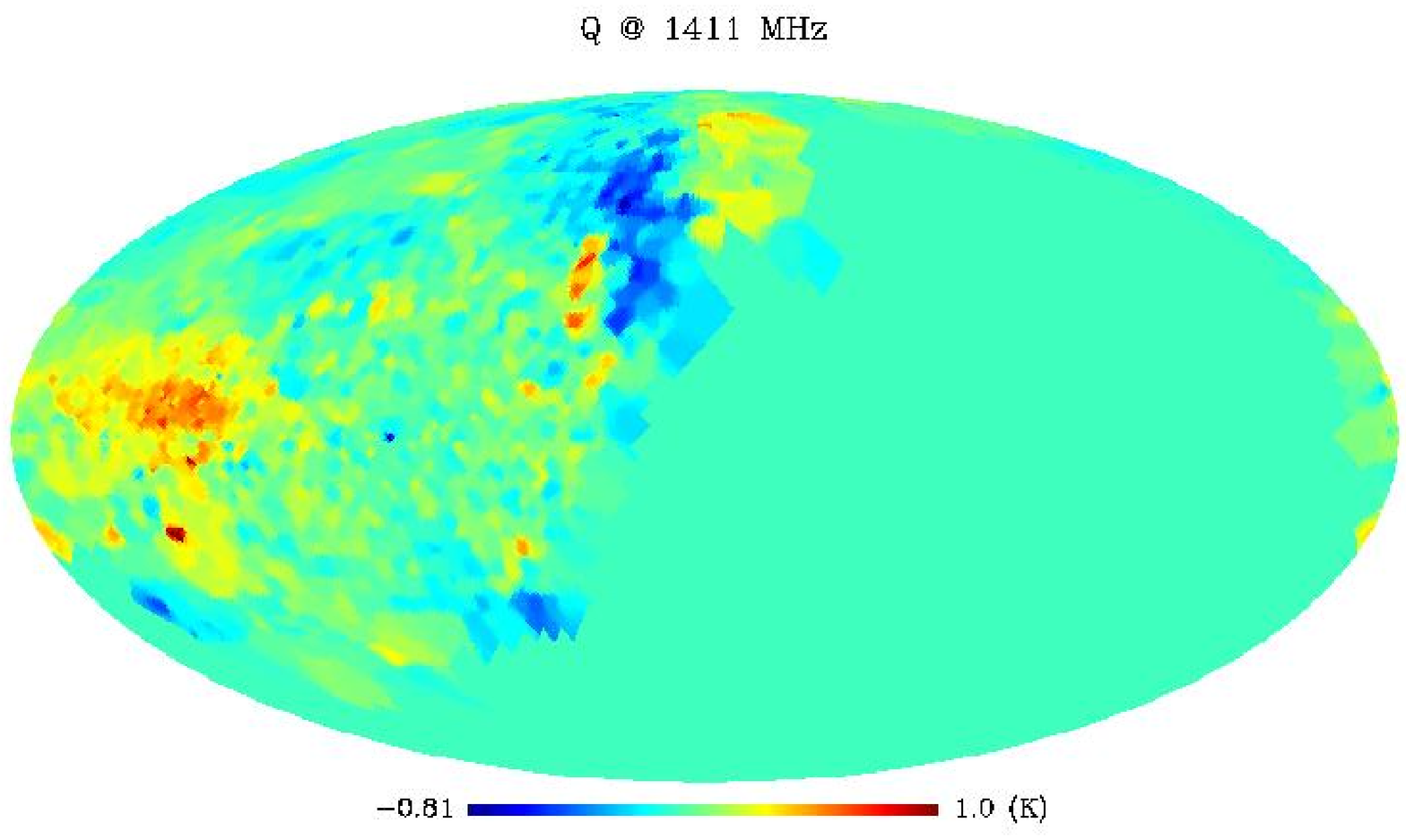}&
   \includegraphics[width=3cm,angle=90,clip=]{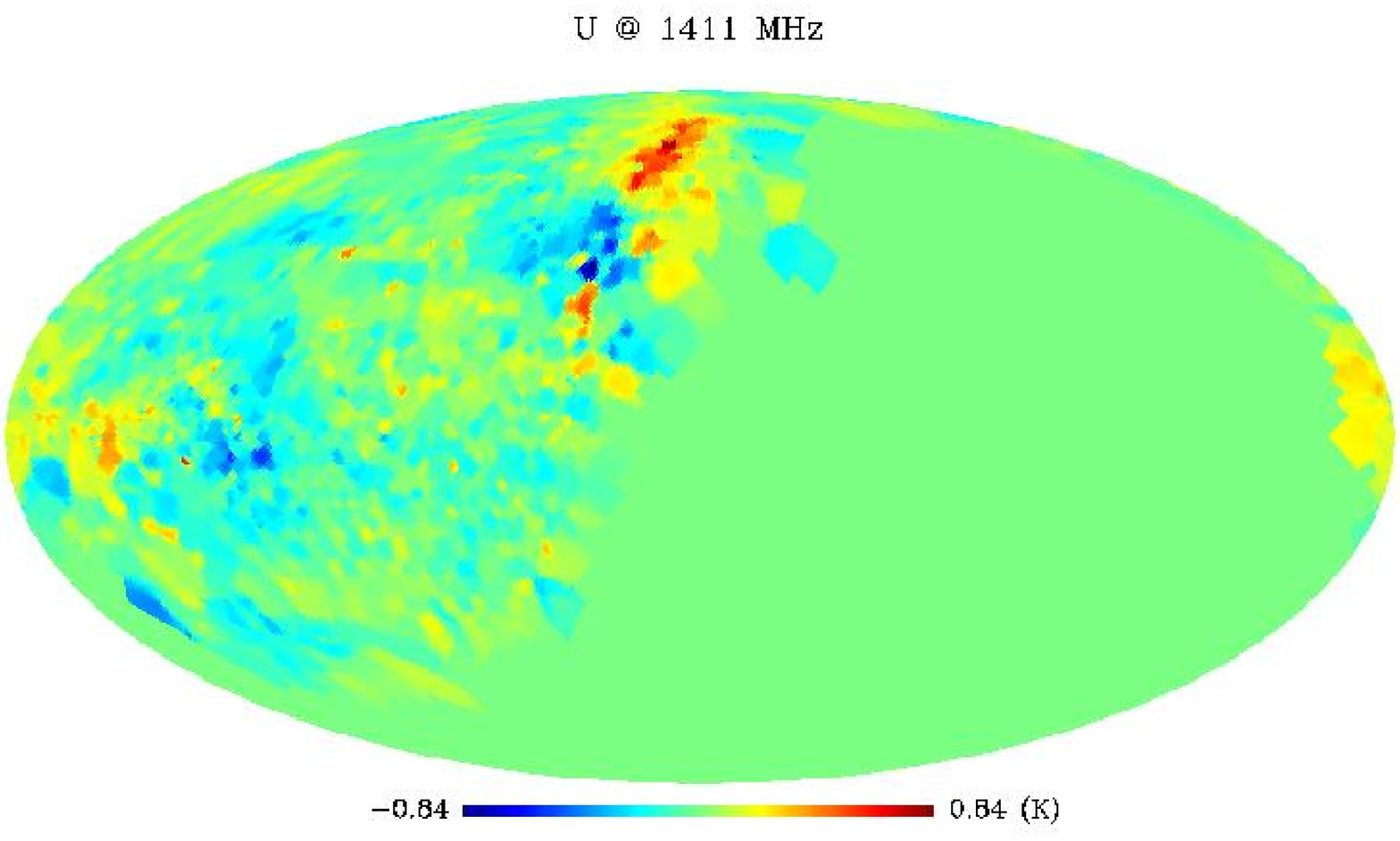}\\
   \end{tabular}
   \caption{$PI$, $Q$, and $U$ maps obtained from the Leiden surveys data
tables using the final version of the interpolation algorithm at all 
frequencies.}
   \label{allmaps}
 \end{figure*}

In Fig.~\ref{allmaps} we show the polarization maps
($PI$, $Q$, and $U$) produced at all frequencies with the final version
of the code described in the previous section.

Using the {\sc Anafast} facility of the HEALPix package, we computed the 
APSs  of the polarized intensity and of the $E$ and $B$ modes for the 
half-sky coverage and for the patches~\footnote{
The monopole of each map has been subtracted before APS estimation 
and the obtained $C_{\ell}$ have been 
divided by the fractional coverage of the considered area
to renormalize it to the whole sky case.}
at the five frequencies of the 
surveys.
In addition to the three better sampled areas we considered 
another region (patch 4) located at 
($70^\circ \leq l \leq 120^\circ$, $-45^\circ \leq b \leq -15^\circ$).
This patch has a sky sampling in the average 
(its APS is then statistically relevant only for $\ell \lsim 50$)
and is characterized by a rather low signal.

\begin{figure*}
\centering
\includegraphics[width=5.5cm,angle=90]{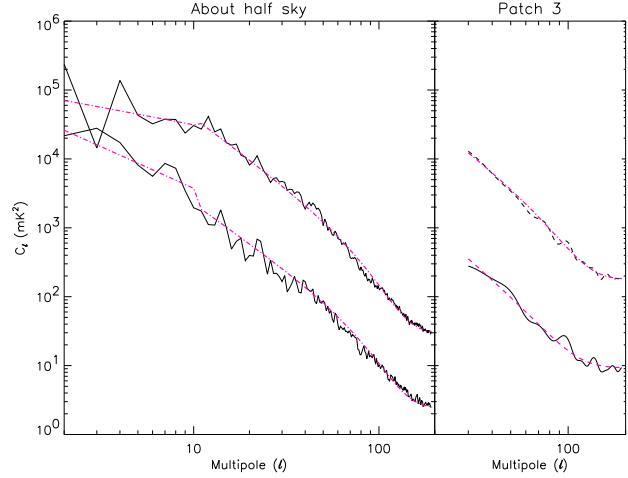}
\vskip 0.5cm
\caption{Polarization APSs at 610~MHz (solid lines) for the survey 
full coverage (left panel) and patch 3 (right panel) together with the 
corresponding best fit curves (dot-dash lines). 
The lower curves in each panel are $C_{\ell}^{PI}$, 
while the other ones represent $(C_{\ell}^{E}+C_{\ell}^{B})/2$
multiplied by 10.}
\label{showfit610}
\end{figure*}

The $E$ and $B$ modes turned out to be extremely 
similar, so that, instead of studying each of them
singularly, we consider their average
$(C_{\ell}^{E}+C_{\ell}^{B})/2$.
In some cases (namely for patch~1 and 2 at 1411~MHz and 
patch~2 and 4 at 408~MHz) 
we could even consider a unique APS defined as 
$(C_{\ell}^{PI}+C_{\ell}^{E}+C_{\ell}^{E})/3$.\\
The maps represent the Galactic polarized synchrotron emission, smoothed 
with the beam of the radiotelescope and contaminated by the noise.
As a consequence, their APSs can be fit as sum of two components,
$C_{\ell} = C_{\ell}^{synch} W_{\ell} + C_{\ell}^{\rm N}$.
We exploit the power law approximation 
$C_{\ell}^{synch} \simeq \kappa \cdot \ell^{~\alpha} \, $ and
assume a symmetric, Gaussian beam, i.e.  
a window function
$W_{\ell}={\rm e}^{-(\sigma_{b} \ell)^2}$,
where 
$\sigma_b= { \theta_{HPBW} ({\rm rad}) / {\sqrt{8 {\rm ln} 2}} }$~.
Under the hypothesis of uncorrelated Gaussian random noise~\footnote{Even in 
the ideal case of negligible systematic effects in the data,
the averaging and the interpolation over the 
poorly sampled data may imply deviations from this simple assumption.
Moreover, the noise level in the maps will be 
connected not only to the instrumental noise, but 
also to the sampling. A posteriori, the APS flattening 
found at high $\ell$ justifies our simple approximation.}
(white noise),
$C_{\ell}^{\rm N} = C_{\ell}^{\rm WN}\sim$~const~.
We have performed the fit on the multipole range $[30,200]$ for 
the patches and $[2,200]$ for the full-coverage maps.
In fact, 
the flattening occurring at higher multipoles helps
recover the noise constant, in spite of the fact that
no reliable astrophysical information 
can be derived for ${\ell} > 100$ from the APSs even in the three better 
sampled regions (\cite{buriganalaporta02}).
For the patches, we found that a single set of the parameters
$\kappa$, $\alpha$ allows us to describe the APS in the entire
range of multipoles. In the case of the survey full 
coverage we needed two sets of $\kappa$, $\alpha$, each of them
appropriate to a certain multipole range.
At all frequencies a change in the APS
slope occurs corresponding to ${\ell} \sim 10$ for both  
$C_{\ell}^{PI}$ and $(C_{\ell}^{E}+C_{\ell}^{B})/2$.\\
As an example, we display in Fig.~\ref{showfit610} the APS and 
the corresponding best fit curve at 610~MHz for the survey 
full coverage and one patch.
%
%
\begin{figure*}
\centering
\includegraphics[width=6cm,angle=90]{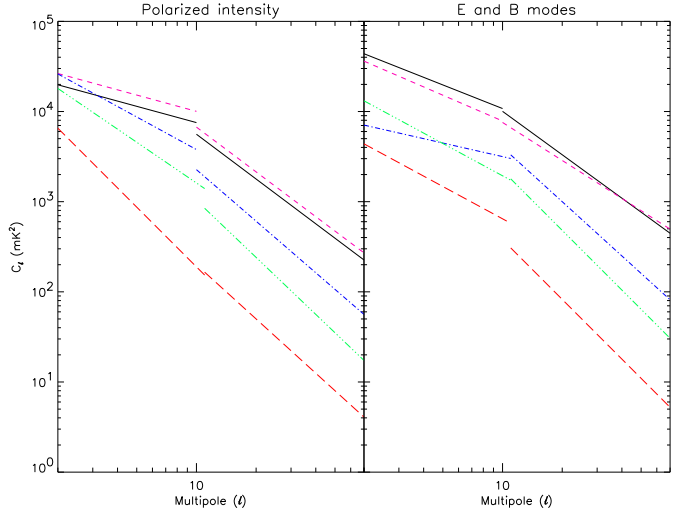}
\vskip 0.5cm
\caption{Synchrotron component of the APS derived for the
survey full coverage for the polarized intensity 
(left panel) and for $(C_{\ell}^{E}+C_{\ell}^{B})/2$ (right panel) 
at the various frequencies: 408 (solid line),
465 (dashes), 610 (dot-dash), 820 (three dot-dash), and 
1411~MHz (long dashes).}
\label{apsplusfit_half}
\end{figure*}
%
\begin{figure*}
\centering
\includegraphics[width=7cm,angle=90]{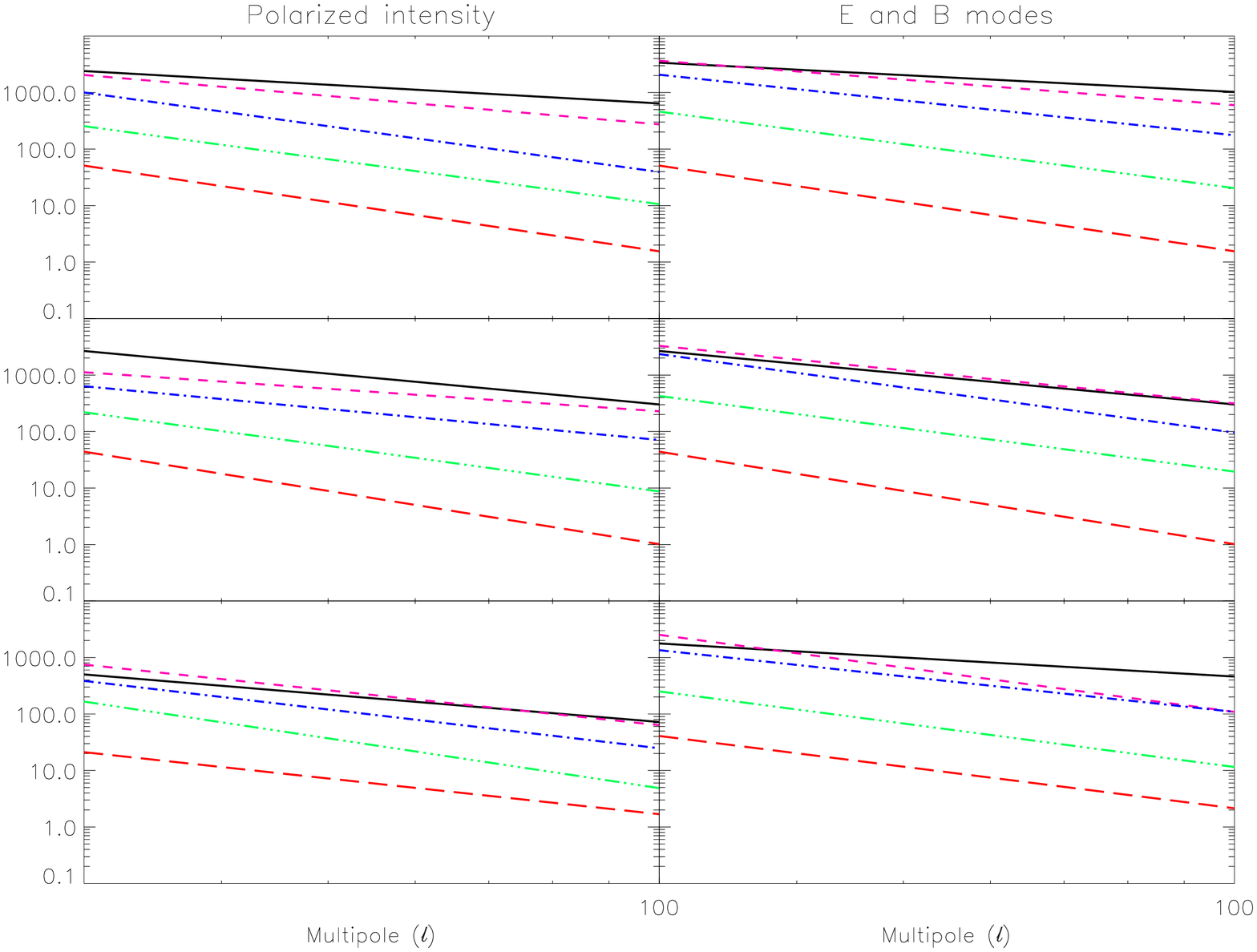}
\vskip 0.5cm
\caption{As in Fig.~\ref{apsplusfit_half} but for 
the selected patches: patch 1 (top panels), patch 2 (middle panels),
and patch 3 (bottom panels).}
\label{apsplusfit_patch}
\end{figure*}
We show in Fig.~\ref{apsplusfit_half} (resp. Fig~\ref{apsplusfit_patch}) 
the recovered synchrotron term, $C_{\ell}^{synch}$,
and list in Table~\ref{BestFitPar_half} (resp. Table~\ref{BestFitPar_patch}) 
the corresponding $\kappa$, $\alpha$ obtained for the survey
full coverage (resp. for the patches).
In the multipole range 
$30 \lsim {\ell} \lsim 100$ 
we found a general trend: from the lowest to the highest frequency
the slope increases from $\sim -$(1-1.5) to $\sim -$(2-3), 
with a weak dependence on the considered sky region.
\begin{table*}[!ht]
\begin{center}
\begin{tabular}{|c|c|c|c|c|c|c|} 
\hline
 $\nu$ & $C_\ell$ & $\kappa_1$ & $\alpha_1$ & $\ell$ & $\kappa_2$ & 
$\alpha_2$ 
\\
 (MHz) & (mK$^2$) & (mK$^2$) & & &  (mK$^2$) &  \\
\hline
 408 & $EB$ & $8.00\cdot10^4$ & -0.87 & 10.86 & $4.00\cdot10^5$ & -1.60 \\
     & $PI$ & $3.00\cdot10^4$ & -0.60 & 10.80 & $2.50\cdot10^5$ & -1.65 \\
\hline
465 &  $EB$ & $7.00\cdot10^4$ & -0.95 & 10.00 & $1.90\cdot10^5$ & -1.40 \\
    & $PI$ & $4.00 \cdot10^4$ & -0.60 & 10.85 & $3.00\cdot10^5$ & -1.65 \\
\hline
610 & $EB$ & $1.00\cdot10^4$ & -0.50 & 10.00 & $4.00\cdot10^5$ & -2.00 \\
    & $PI$ & $6.00\cdot10^4$ & -1.20 & 10.85 & $1.8\cdot10^5$ & -1.90 \\
\hline
820 & $EB$ & $3.00\cdot10^4$ & -1.19 & 11.00 & $3.5\cdot10^5$ & -2.20 \\
    & $PI$ & $5.10\cdot10^4$ & -1.50 & 11.00 & $1.30\cdot10^5$ & -2.10 \\
\hline
1411 & $EB$ & $1.00\cdot10^4$ & -1.19 & 11.00 & $6.00\cdot10^4$ & -2.20 \\
     & $PI$ & $3.00\cdot10^4$ & -2.20 & 11.00 & $2.00\cdot10^4$ & -2.00 \\
\hline
\end{tabular}
\end{center}
\caption{Values of the best fit parameters $\kappa$ and $\alpha$
obtained in the analysis of the survey full coverage APS at all the  
frequencies (as shown in Fig.\ref{apsplusfit_half}). 
Relative errors on the best fit parameters are $\sim 10\%$.
Two multipole ranges are considered: $[2,\ell]$ and $[\ell,200]$
($\ell$ was also a free parameter in the fit).
$EB$ refers to $(C_{\ell}^{E}+C_{\ell}^{B})/2$.
}
\label{BestFitPar_half}
\end{table*}
\begin{table*}[!ht]
\begin{center}
\begin{tabular}{|c|c|c|c|c|c|c|c|c|c|c|c|}
\hline
       &&\multicolumn{2}{c|}{408~MHz}& \multicolumn{2}{c|}{465~MHz} 
       &\multicolumn{2}{c|}{610~MHz} &\multicolumn{2}{c|}{820~MHz} 
       &\multicolumn{2}{c|}{1411~MHz} \\   
      & & $\kappa$~(mK$^2$)& $\alpha$ & $\kappa$~(mK$^2$) & $\alpha$ & 
$\kappa$~(mK$^2$) 
& $\alpha$ & 
       $\kappa$~(mK$^2$) & $\alpha$ & $\kappa$~(mK$^2$) & $\alpha$ \\
\hline
  P~1 & $EB$ & $9.80\cdot10^4$ & -0.99 & $6.00\cdot10^5$ & -1.50 &
            $2.21\cdot10^6$ & -2.05 & $3.08\cdot10^6$& -2.59 &
            $9.80\cdot10^5$& -2.90 \\
          & $PI$ & $1.00\cdot10^5$ & -1.10 & $6.00\cdot10^5$ & -1.67 & 
            $9.44\cdot10^6$ & -2.69 & $2.02\cdot10^6$ & -2.64 &
            $9.80\cdot10^5$  & -2.90\\
\hline
  P~2 & $EB$ &  $1.26\cdot10^6$ & -1.81 &  $2.50\cdot10^6$ & -1.95 
            & $2.00\cdot10^7$& -2.66 & $2.58\cdot10^6$& -2.56  
            & $1.85\cdot10^6$& -3.13    \\
          & $PI$ & $1.26\cdot10^6$ & -1.10 & $1.00\cdot10^5$ & -1.32 & 
               $3.10\cdot10^5$ & -1.82 & $2.00\cdot10^6$ & -2.68 &
               $1.85\cdot10^6$  & -3.13  \\
\hline
  P~3 &  $EB$ & $8.00\cdot10^4$  & -1.12 & $1.81\cdot10^7$ & -2.61 
           & $1.65\cdot10^6$ & -2.09 & $1.52\cdot10^6$& -2.56 
           & $1.70\cdot10^5$ & -2.45     \\
          & $PI$ & $1.20\cdot10^5$ & -1.61 & $8.00\cdot10^5$ & -1.32& 
              $9.00\cdot10^5$ & -2.28 & $3.53\cdot10^6$ & -2.93 &
              $2.70\cdot10^5$ & -2.10 \\
\hline
  P~4 &  $EB$ & $2.00\cdot10^4$ & -0.91 & $2.60\cdot10^4$ & -1.08 
           & $3.10\cdot10^6$ & -2.69 & $3.36\cdot10^6$& -3.07 
           & $6.10\cdot10^5$ & -2.97     \\
          & $PI$ & $2.00\cdot10^4$ & -0.91 & $1.10\cdot10^5$ & -1.65 &
            $2.60\cdot10^5$ & -2.24 & $7.00\cdot10^4$ & -2.28 &
            $1.00\cdot10^6$ & -3.33 \\
\hline
\end{tabular}
\end{center}
\caption{Values of best fit parameters $\kappa$ and $\alpha$
obtained in the analysis of the APS of the patches at all the  
frequencies (as shown in Fig.\ref{apsplusfit_patch}).
Relative errors on the best fit parameters are $\sim 10\%$. 
The considered multipole range is [30-200].
$EB$ refers to $(C_{\ell}^{E}+C_{\ell}^{B})/2$.}
\label{BestFitPar_patch}
\end{table*}

\section{Comparison with the DRAO polarization survey}

\begin{figure*}
\centering
\includegraphics[width=8cm,angle=90]{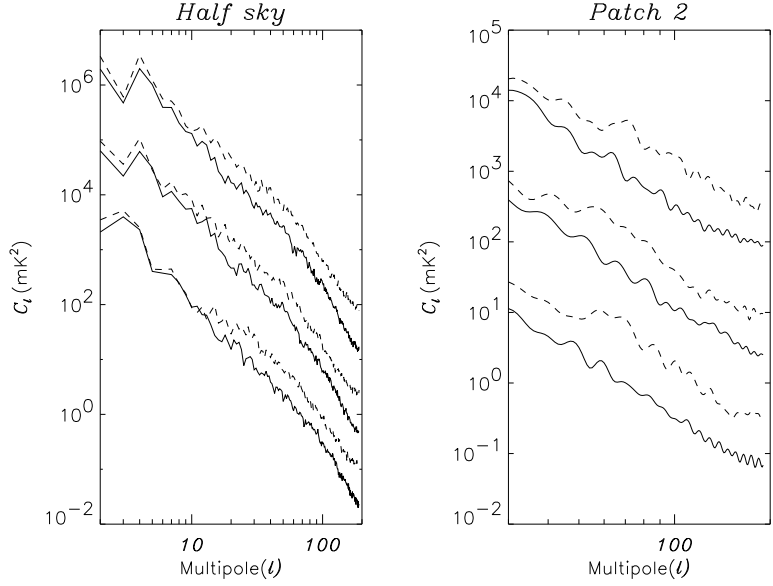}
\caption{Comparison between the DRAO APSs (solid lines) 
and the APSs of the maps reconstructed by our code (dashes) 
from the 1411~MHz Leiden survey
properly rescaled according to Wolleben~(2005). 
In each panel, the three pairs of curves 
(from the bottom to the top) correspond to the $PI$, $E$, and $B$ modes;
the $E$ and $B$ APSs have been multiplied 
respectively by 20 and 700.} 
\label{DRAOvsSpoelstra}
\end{figure*}

Soon, a new all-sky survey at 1.4~GHz will become available. 
It results from the merging of two different polarization surveys 
having the same angular resolution of about $36'$: the Northern part 
of the sky has been observed with the DRAO 26~m telescope 
(\cite{wolleben03}), for the Southern hemisphere 
the 30~m Villa Elisa telescope was used 
(\cite{testori03}).
The first was meant to compensate for the lack of information typical of 
the Leiden surveys and it improves their 
sampling and sensitivity.
However the Leiden surveys remain so far a unique tool 
for absolute calibration in the Northern sky, therefore
the DRAO data have been tightened to them. 
A second 
observing period at DRAO has recently been completed. 
The final map is fully sampled in right ascension
and has a spacing in declination ranging 
between $0.25^{\circ}$ and $2.5^{\circ}$ (\cite{wolleben05}). 
In Fig.~\ref{DRAOvsSpoelstra} we compare the APS  
of $C_{\ell}^{PI},~C_{\ell}^{E}$, and $C_{\ell}^{B}$ 
derived from  a preliminary version of the DRAO survey  
with the ones of the maps we constructed using the Leiden survey 
at 1411~MHz. 
For the full coverage maps the APSs almost coincide
at $\ell \lsim 10$. At $\ell \gsim {\rm few} \times 10$
the slopes are in good agreement in all cases,
while the amplitudes derived from the Leiden survey are 
larger by a factor $\sim 2.5$ for the full-coverage maps 
and $\sim 4$ for the patches. As a consequence, the fitted amplitudes
reported in the previous section might translate in an overestimate
of the synchrotron emission APS at 1.4 GHz. We checked that the signal 
in the DRAO survey is typically lower than in map we constructed from the 
Leiden survey at 1411 MHz, which seems consistent with the amplitude
discrepancy found. A detailed analysis of the final version of the recent 1.4~GHz 
polarization data and its comparison with the Leiden 1.4~GHz survey
is in progress.  
We stress here the good agreement between the slopes of the APSs 
obtained from the two data sets that confirms the steepness of 
the APS we found at 1.4~GHz from the Leiden survey.

\section{Multifrequency analysis}

In Sect.~5 we studied the APS of the
Galactic diffuse polarized emission at each frequency, 
focusing on its multipole dependence.
We first consider here the behaviour of the synchrotron APS 
at some fixed values of the multipole as function of frequency,
being aware of the depolarization phenomena.
In the second subsection we provide a qualitative 
interpretation of the APS steepening with increasing frequency
in terms of Faraday depolarization effects.

\subsection{Frequency dependence of the APS amplitude 
and Faraday depolarization}

The theoretical intrinsic behaviour of the synchrotron total intensity
emission predicts a power law dependence of the APS amplitude
on frequency:
$$T^{synch}\propto \nu^{-\beta-2} \Rightarrow C_{\ell}^{synch}(\nu) \propto \nu^{2(-\beta-2)}\; .$$
The values of $\beta$ deduced from radio observations change
with the considered sky position and range between 
$\sim 0.5$ and $\sim 1$ (see Reich \& Reich~1988 and Platania et al.~1998).
In principle, one would expect the synchrotron emission to be highly 
polarized, up to a maximum of $\sim 75\%$ (see \cite{ginzburg65}).
In the ideal case of constant degree of polarization
the above formulae would apply also to the
polarized component of the synchrotron emission, once the brightness 
temperature has been properly rescaled.
However, depolarization effects are relevant at radio frequencies, 
so that the predicted correlation between total and polarized intensity
do not show up at all (except for the NPS and the Fan Region
that are clearly evident
both in total intensity and 
polarization maps at about 1.4~GHz).
The current knowledge of $\beta$ for the Galactic diffuse 
polarized emission is very poor, but already indicates  
that due to depolarization phenomena the observed $\beta$ 
can be much lower than $\sim0.7$ (e.g. for the NCP 
Vinyajkin \& Razin 2002 quote $\beta\sim -0.13$).

\begin{figure*}
\centering
\includegraphics[width=10cm]{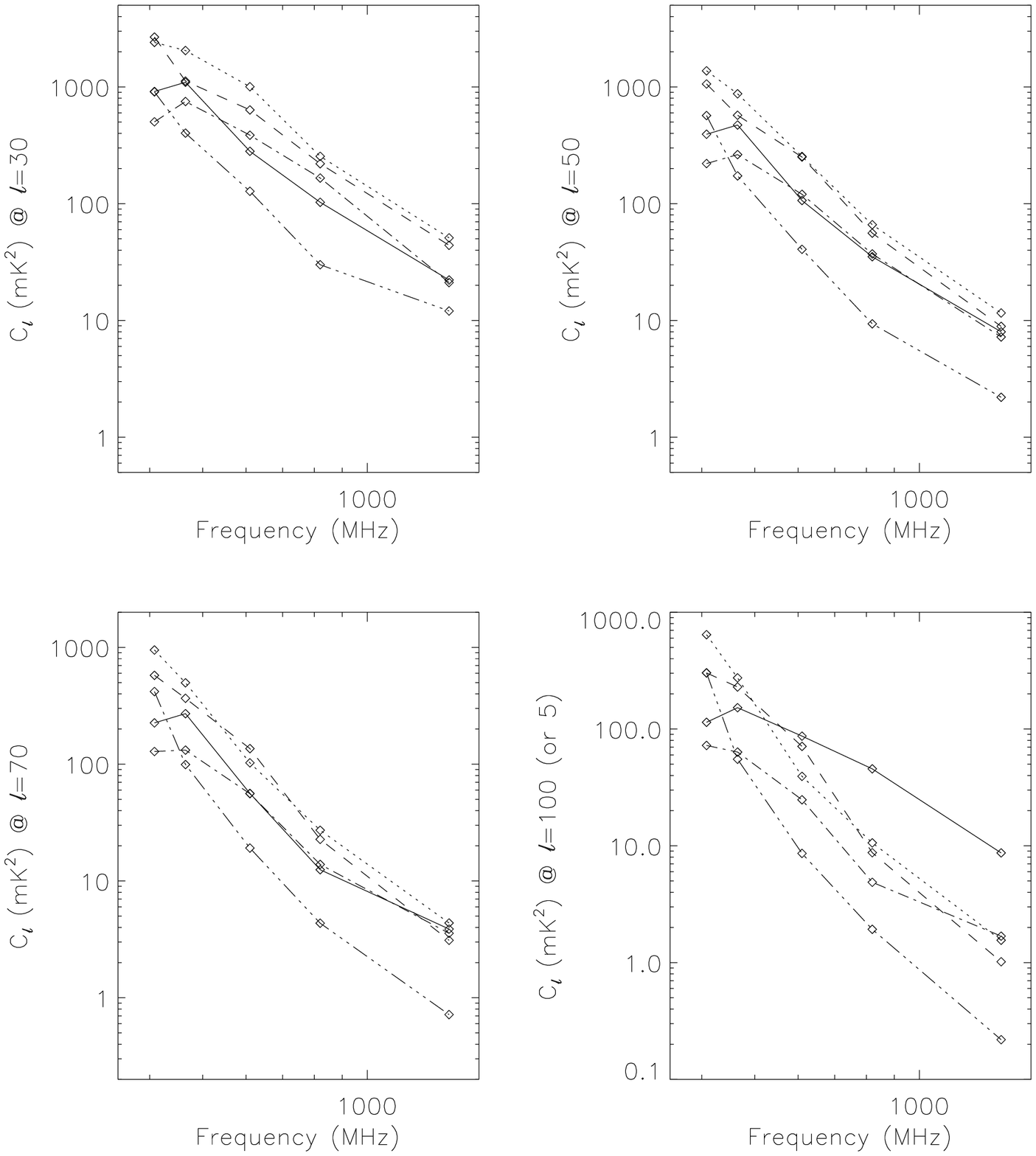}
\caption{Synchrotron $PI$ APS at all frequencies for some 
fixed representative multipoles, $\bar{\ell}$ close to the boundary 
and in the middle of the range of validity of our analysis.
We considered $\ell=30, 50, 70, 100$ for the patches and 
$\ell = 5, 30, 50, 70$ for the full coverage maps.
The diamonds represent the $C_{\bar{\ell}}^{synch}(\nu)$
at the survey frequencies. 
Different lines correspond to different sky regions,
namely: 
full survey coverages (solid line), 
patch~1 (dots), patch~2 (dashes), 
patch~3 (dot-dashes), patch~4 (three dots-dashes).} 
\label{clfreq}
\end{figure*}

In Fig.~\ref{clfreq} we plot $C_{\ell=\tilde{\ell}}^{synch}(\nu)$ at 
boundary and intermediate values of the statistically significant 
multipole range in each coverage case. For the patches we chose
$\tilde{\ell} = 30, 50, 70, 100$, whereas for the half-sky maps we 
considered $\tilde{\ell} = 5, 30, 50, 70$. 
The APS amplitude actually decreases with frequency,
but our results are not well fitted by a power law and exhibit
shapes significantly flatter than that of the observed synchrotron
emission in total intensity (approximately 
$C_{\ell}^{synch}(\nu) \propto \nu^{-5.4}$), as expected in
the presence of frequency-dependent depolarization effects.

An electromagnetic wave travelling through a magnetized plasma 
will undergo a change in the polarization status. 
In particular its polarization vector will be rotated by an angle
$$\Delta\phi{\rm [rad]} = RM{\rm [rad/m}^2]\cdot \lambda^{2} {\rm [m}^2] \; .$$
Here the rotation measure, $RM$, is the line of sight integral 
$$RM [{\rm rad/m}^2]= 0.81 \int n_{e}[{\rm cm}^{-3}] 
\cdot B_{\Vert}[\mu{\rm G}] dl[{\rm pc}]  \; ,$$ 
where $n_{e}$ is the electron density and $B_{\Vert}$
is the component of magnetic field along the line of sight.

\begin{figure*}
\centering
\includegraphics[width=10cm]{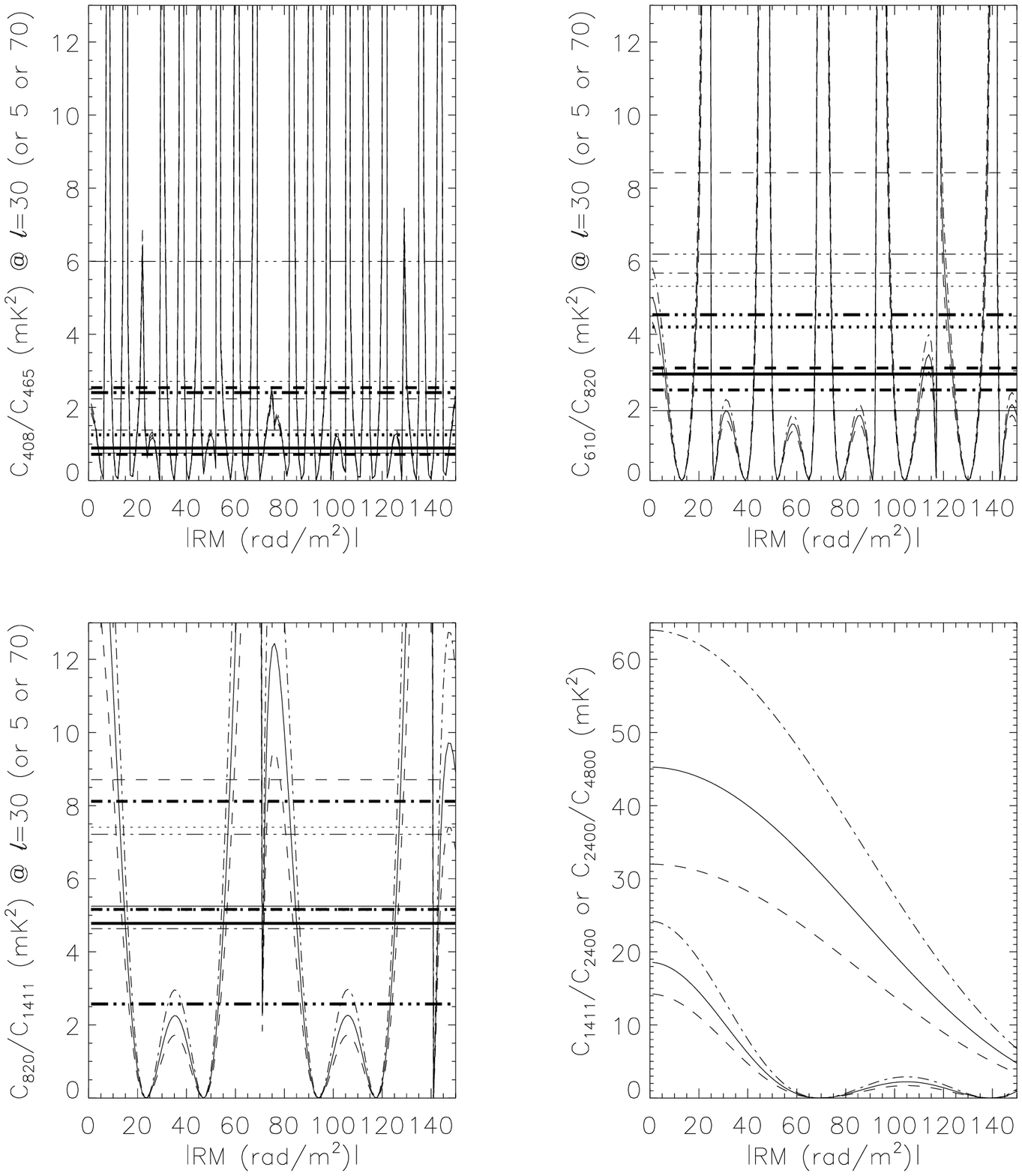}
\caption{Ratios 
($C_{\tilde{\ell}}^{synch}(\nu_{1})/C_{\tilde{\ell}}^{synch}(\nu_{2})$)
between the polarized synchrotron APSs at two frequencies for 
a pair of representative multipoles
in the range of validity of our analysis.
For the patches we plot the ratio 
at $\ell$~=~30 and 70, 
whereas for the full coverage maps we chose $\ell$~=~5 and 30.
The horizontal lines represent the observed ratio 
corrected for beamwidth depolarization; thick lines 
refer to the lower of the two considered multipoles.
Different sky regions are considered:
full survey coverages (solid line), 
patch~1 (dots), patch~2 (dashes), 
patch~3 (dot-dashes), patch~4 (three dots-dashes).
The oscillating curves represent the theoretical ratio as a function
of $RM$ in case of pure Faraday depolarization 
for three choices of the synchrotron emission frequency spectral 
index: $-(\beta+2) = -2.5$ (dashes), solid $-(\beta+2) = -2.75$ (solid 
lines), and $-(\beta+2) = -3$ (dot-dashes) . 
The last panel  show the same three curves in 
the hypothetical case of availability 
of observations at higher frequencies. } 
\label{freqratio}
\end{figure*}

Given $\Delta\phi \propto \nu^{-2}$, the lower the frequency the 
greater the importance of the Faraday rotation effect. 
Approximating the interstellar medium (ISM) as a superposition of 
infinitesimal layers each characterized by constant 
$n_{e}$ and $B_{||}$ ({\it slab model}, see \cite{burn66}), 
the observed polarization intensity relates to the intrinsic 
one according to the expression:
$$T^{p,obs}(\lambda)=T^{p,intr}(\lambda) \cdot 
|{\sin\Delta\phi}/{\Delta\phi}| \; .$$
Therefore the net effect of
a differential rotation of the polarization angle
is a depletion of 
the polarized intensity, commonly referred to as Faraday depolarization.
Given $\left({T^{p}}\right)^2 \propto C_{\ell}(\nu) $ the above 
formula reads:
$$C_{\ell}(\nu)=C_{\ell}^{intr}(\nu) \cdot 
({\sin{\Delta\phi}}/{\Delta\phi})^2 \; ,$$
where $C_{\ell}^{intr}(\nu) = 
\kappa \cdot{\ell}^{\alpha}\cdot\nu^{2(-\beta-2)} \, .$
At a given multipole, 
the ratio between the APSs at two frequencies can be written as:
$${ C_{\ell}(\nu_1) }/{ C_{\ell}(\nu_2) }=({\nu_1}/{\nu_2})^{-2\beta-4} 
\cdot ({\Delta\phi_2} / {\Delta\phi_1}) ^2 \cdot 
({\sin{\Delta\phi_1} / \sin{\Delta\phi_2}})^2 \; .$$
Given a certain value of $\beta$, the expression on the right side of
this equation can be displayed as a function of $RM$ (see 
Fig.~\ref{freqratio}). 
The $RM$ intervals in which the curve (obtained for a certain $\beta$) 
intersects the constant line representing the observed value of this ratio 
provide information about the possible $RM$ values\footnote{Clearly, the 
$RM$ values identified with this method do not refer to a precise 
direction in the sky, but to the considered sky area.}. This holds the in
case of pure Faraday depolarization. In reality there are at least two 
other phenomena affecting the observed polarized intensity, the beamwidth 
and the bandwidth depolarization 
(see \cite{GardnerWhiteoak66} and \cite{sokoloff}).  
If the polarization angles vary within the beam area, then observations 
of the same sky region with different angular resolution will give 
an apparent change of the polarized
intensity  with frequency ({\it beamwidth depolarization}). 
Suppose $\theta_{HPBW,1}$ and 
$\theta_{HPBW,2} (< \theta_{HPBW,1})$
are the angular resolutions respectively at $\nu_1$ and $\nu_2$.
The effect can be removed by smoothing the map at the 
frequency $\nu_2$ to the (lower) angular resolution of the 
map at the frequency $\nu_1$ or, equivalently, multiplying 
the observed APSs ratio, ${C_{\ell}(\nu_{1})}/{C_{\ell}(\nu_{2})}$ 
by ${\rm e}^{(\Delta\sigma_{b}\ell)^2}$, 
where $\Delta\sigma_{b}={\sigma_1}^2-{\sigma_2}^2$  
($\sigma_i=\theta_{HPBW,i}$~(rad)/${\sqrt{8 {\rm ln} 2}}\;$; $i=1,2$). 
This correction~\footnote{Once the 1411~MHz map is properly smoothed 
to the lower resolution of the 820~MHz 
survey, the bulk of possible differencies due to 
beam depolarization effects is eliminated
and no longer contributes to the value of the ratio between the APSs 
at the two frequencies.   
Consequently, beam depolarization is not an issue in the following 
analysis.} 
has been applied to convert the APSs shown 
in Fig.~\ref{clfreq} to the ratios shown in Fig.~\ref{freqratio}.\\
Another depolarization effect is introduced by the 
receiver bandwidth: the polarization angle will change by an amount
$$\Delta\psi = - RM \cdot {\lambda_{0}}^2 \cdot {2\Delta\nu}/{\nu_{0}} \; ,$$
where $\lambda_{0}$ and $\nu_{0}$ are the values at the bandwidth 
centre. 
The polarization of the incoming radiation will be reduced by a factor
${\sin\Delta\psi}/{\Delta\psi}$, that in our case is negligible
(${\Delta\nu}=1.7$~MHz at the two lower frequencies, 
${\Delta\nu}=3.6$~MHz at the intermediate ones and 
${\Delta\nu}=7.2$~MHz at the highest one).

In Fig.~\ref{freqratio} we plot the 
$C_{\tilde{\ell}}^{synch}(\nu_{1})/C_{\tilde{\ell}}^{synch}(\nu_{2})$  
at $\tilde{\ell} = 30, 70$ (resp. at $\tilde{\ell} = 5, 30$)
in the case of the patches 
(resp. in the case of the full-coverage maps).  

The information coming from the lower frequencies is too poor
to be able to learn something about $RM$s; on the other hand, the ratio 
between the APSs at 820 and 1411~MHz give some indications.
For a reasonable range of synchrotron emission
frequency spectral indices, the following values are compatible with 
the data (see left-bottom panel of Fig.~\ref{freqratio}): the relatively spread 
intervals 9-17, 53-60, 75-87, 123-130~rad/m$^2$ together with two narrow 
intervals at 70 and 140~rad/m$^2$. In general there is a rather good
agreement among the results coming from the patches and the full-coverage 
map. Spoelstra~(1984) reported as a typical value $RM = 8$~rad/m$^2$, 
however this is likely a lower limit to the real values
(in addition, beam depolarization effects were not taken into account in 
that study, likely implying a certain underestimation of $RM$).

\begin{figure*}
\centering
\includegraphics[width=10cm]{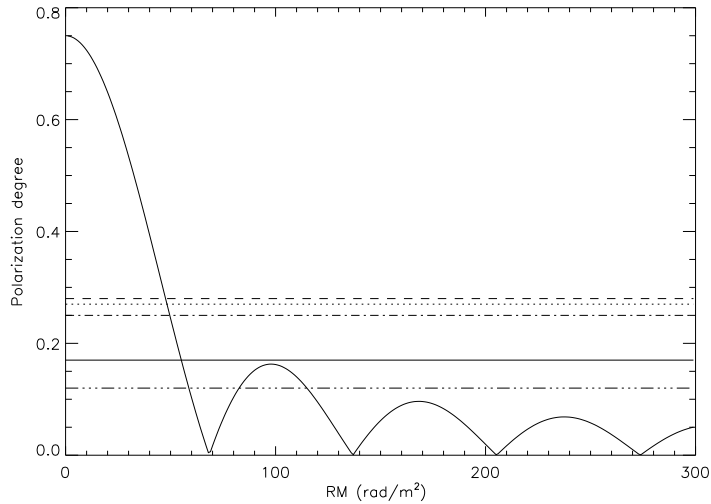}
\caption{The maximum possible value of the observed polarization degree 
(i.e. $\Pi^{obs}=\Pi^{intr} \cdot |\sin\Delta\phi/\Delta\phi|$, where
 $\Pi^{intr}=0.75$) is displayed as function of the $RM$. 
The horizontal lines represent the mean polarization degree 
observed in the considered sky regions. 
Survey full coverage (solid line), 
patch~1 (dots), patch~2 (dashes), 
patch~3 (dot-dash), patch~4 (three dots-dash).}
\label{plotpoldeg}
\end{figure*}

Another constraint to the possible values of the $RMs$
comes from the polarization degree, defined as 
$$\Pi=T^{p}/T\; .$$ 
Exploiting the total intensity northern sky map 
at 1.4~GHz (Reich \& Reich~1988), we estimate the 
mean polarization degree
$\overline{\Pi}$ in the portion 
of the sky covered by the Leiden survey at 1411 MHz.
We find $\overline{\Pi}^{obs}\sim 17, 27, 28, 25, 12\%$ for the full 
coverage and for the patch 1, 2, 3, and 4, respectively. 
By using the preliminary DRAO survey, for the same sky regions 
we obtaine slightly lower values, i.e.
$\overline{\Pi}^{obs}\sim 12, 22, 24, 20,$ and $9\%$, respectively.
Given the relation existing between the intrinsic and
observed brightness polarized temperature in the slab model
approximation, the observed degree of polarization can be written as:
$$\Pi^{obs}=\Pi^{intr} \cdot|\sin\Delta\phi/\Delta\phi|\; .$$
In Fig.~\ref{plotpoldeg} the maximum possible value of $\Pi^{obs}$  
(derived assuming $\Pi^{intr}=0.75$ ) is
displayed as a function of the $RM$. Each horizontal line 
represents the mean polarization degree actually observed for 
the Leiden surveys in the considered coverage cases\footnote{ Note that
the observed mean polarization degree might be due to the combination
of Faraday depolarization and beam depolarization. If we were
able to correct for beam depolarization effects, these lines 
would be shifted upwards in Fig.~\ref{plotpoldeg}, leading to 
smaller values of RMs.
Therefore the results of our analysis are rather conservative 
and provide upper limits to the RMs compatible 
with observations in the considered patches. }. 

Except for patch 4, the observed polarization degree 
sets an upper limit to the $RM$ of $\sim$~50~rad/m$^2$
(having assumed in Fig.~\ref{plotpoldeg} the maximum 
degree of polarization). 
Consequently the lowest of the $RM$ intervals previously 
identified (exploiting the APSs amplitude ratio at the two highest 
frequencies of the Leiden surveys) seems to be the most 
reasonable. 

The above considerations hold under the following 
assumptions:
$i)$ the physical properties of the magneto-ionic ISM responsible for 
depolarization are homogeneous along the line of sight;
$ii)$ the bulk of the observed polarization signal 
comes spatially from the same regions
at all frequencies.
Under significant violations of these 
assumptions the meaning of derived $RM$ is 
by definition unclear. \\ 
The real situation is generally much more complicated, due to the 
existence of turbulence in both the Galactic magnetic field and the 
ISM electron density depolarizing the diffuse synchrotron
emission (see Haverkorn et al.~2004).
Another aspect is the presence of foreground magneto-ionic structures (Faraday screens) 
modulating the background synchrotron diffuse emission 
(see \cite{Wolleben&Reich04} and \cite{Uyaniker03})
, which renders the understanding of the local 
ISM very difficult. There is 
a lack of information about the nature and 
spatial distribution of these structures and, consequently, 
about the corresponding depolarization effects. \\
Nevertheless, in patch 1, 2 and 3
the polarization vectors are extremely well aligned and 
similarly distributed at the two higher frequencies
of the Leiden surveys. 
This observational fact suggests that for these areas the
{\it slab model} is a reasonable approximation of the local ISM,
at least for the angular scales investigated in the present
analysis. Consequently, the above considerations on the $RM$s
 derived from the the APS ratios should be basically correct
 with respect to these regions (whereas they should be taken
 with caution in the other considered cases).

Finally, the right-bottom panel of 
Fig.~\ref{freqratio} shows, as an example,
that a multifrequency sky mapping at $\nu \sim$~some~GHz
would provide direct information on Faraday depolarization, 
significantly reducing the degeneracy 
on $RM$ (unavoidable in maps limited to $\nu \sim 1$~GHz).

\subsection{Effect of the Faraday depolarization on the APS slope}

\begin{figure*}   
   \centering
   \begin{tabular}{cc}
   \includegraphics[width=4cm,angle=90,clip=]{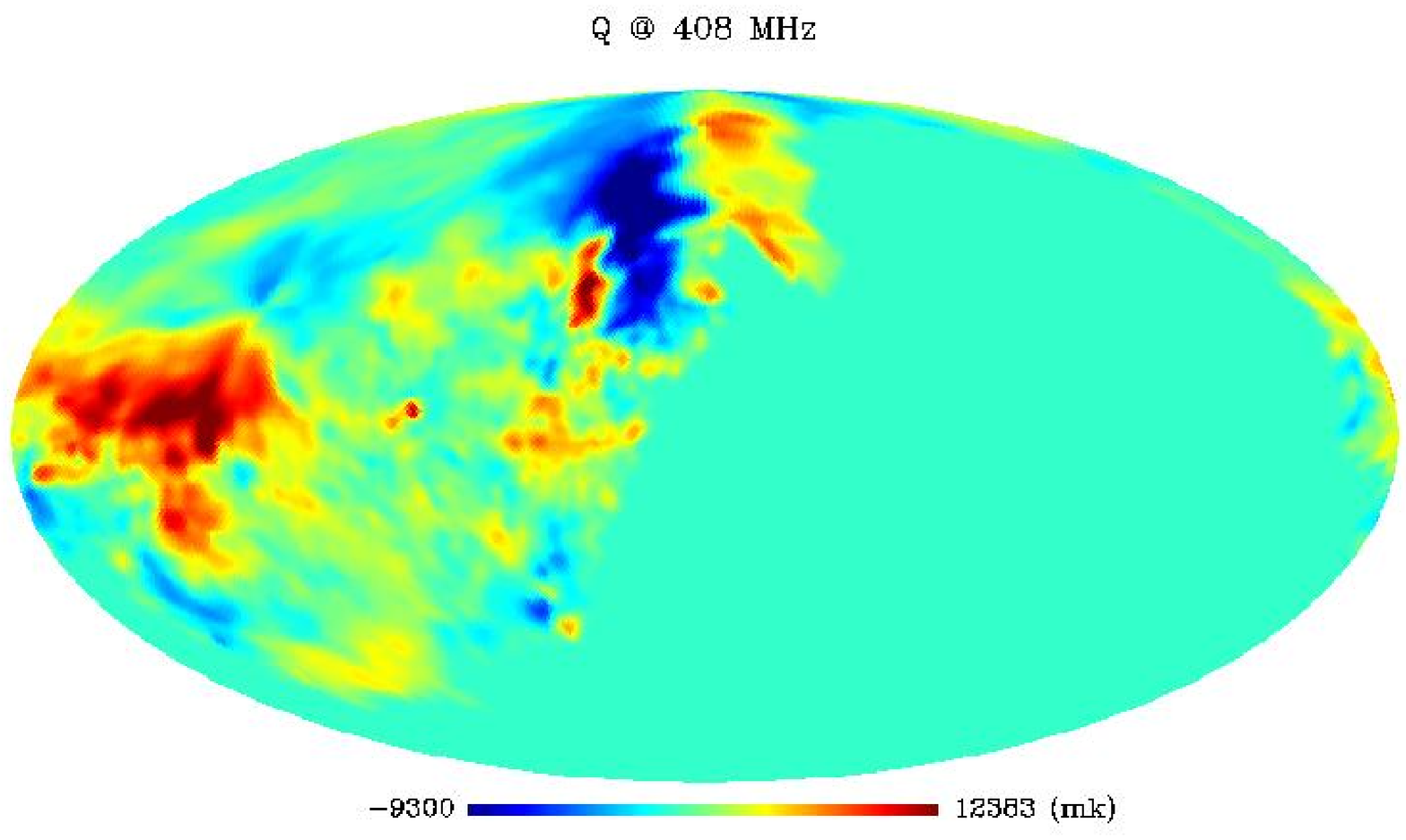}&
   \includegraphics[width=4cm,angle=90,clip=]{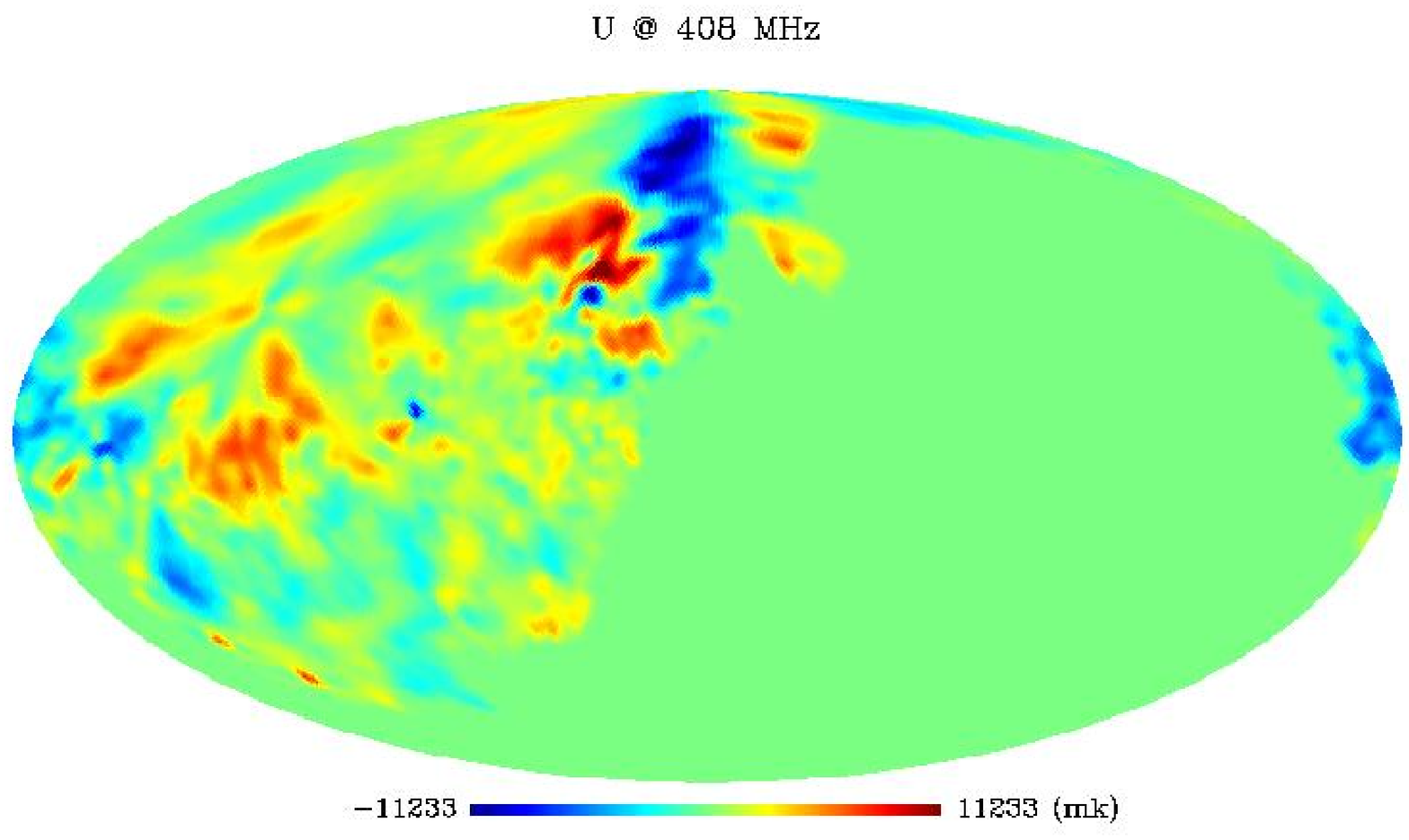}\\
   \multicolumn {2}{c} {\includegraphics[width=4cm,angle=90]{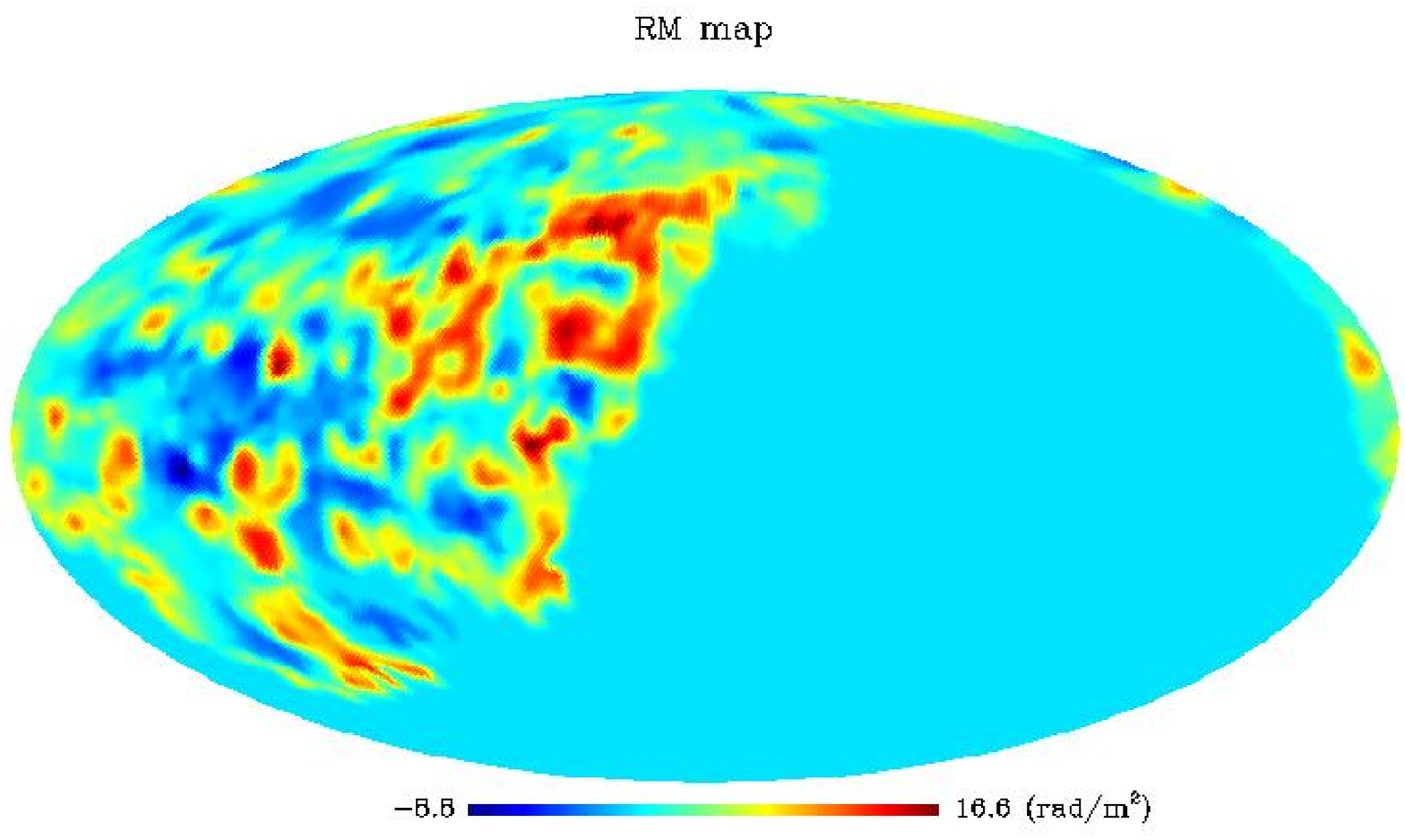}}\\ 
   \includegraphics[width=4cm,angle=90,clip=]{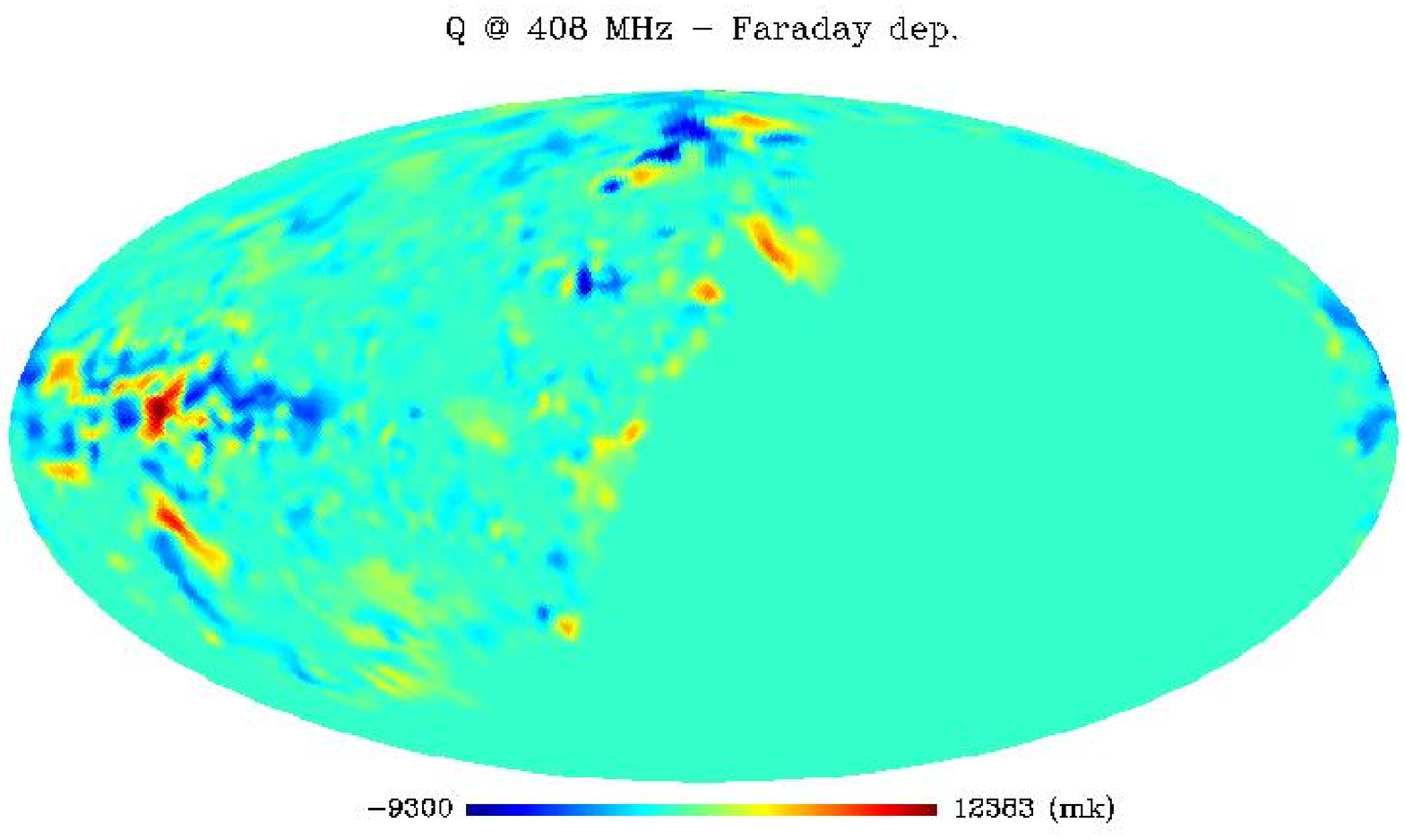} &
   \includegraphics[width=4cm,angle=90,clip=]{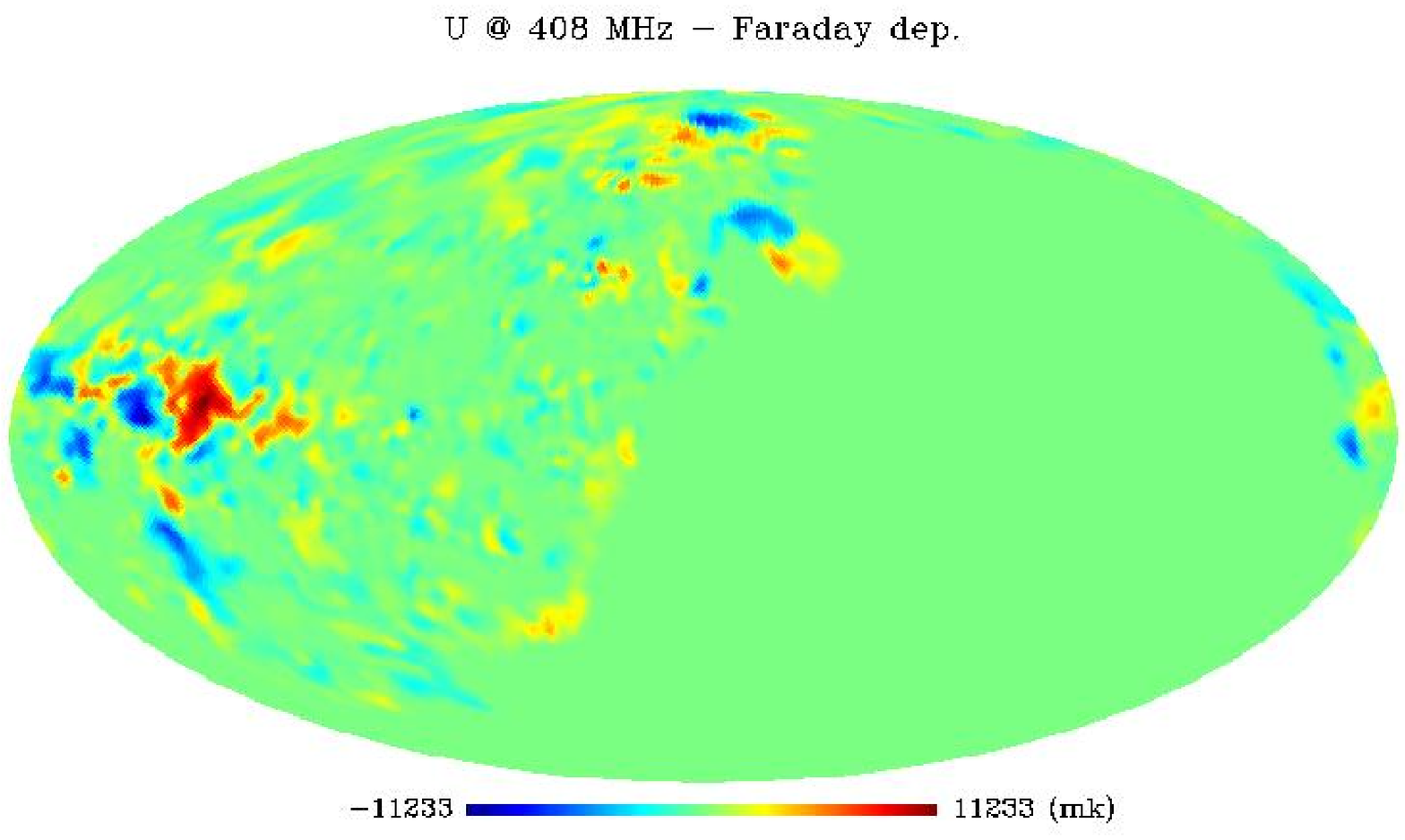}\\
   \end{tabular}
   \caption{Maps of the Stokes parameters before (upper panels) and 
    after (lower panels) the transformation for Faraday depolarization.
    In the middle is placed the $RM$ map used in the simulation.
    All maps have been smoothed to $2.3^{\circ}$. In the depolarized
    maps the large scale structure appear significantly depressed
    respect to the input ones, whereas the small scale structures are
     more relevant.}
   \label{DepMap408}
 \end{figure*}

Our analysis of the Leiden surveys showed that the synchrotron 
emission APS steepens with increasing frequency (see Sect.~ 5).
One possible explanation for such a behaviour of the APS 
 relies on depolarization arguments:
fluctuations of the electron density and/or of the magnetic field (in strength 
and/or direction) might redistribute the 
synchrotron emission power from the large to the small angular 
scales, creating fake structures. 
As a consequence the corresponding APS will become 
flatter, the effect being more important at the lower frequencies.
Sticking to the {\em slab model} to describe the ISM, we
simulated the Faraday depolarization effects on the APS
at 408~MHz and 1420~MHz as follows. 
We assumed that the 1420~MHz polarization map (\cite{wolleben05}) 
represents the intrinsic Galactic synchrotron emission and used 
a spectral index of $-2.75$ to scale it down to 408~MHz\footnote{The
spectral index we used is rather steep for the polarized emission.
However the present exercise is meant to justify the change of 
the APS slope with increasing frequency. It is not intended to 
reproduce the 408~MHz map reconstructed from the Leiden data. 
This would require at least a frequency
spectral index map appropriate to the polarization signal.
The current knowledge of the spectral index distribution 
is limited to the total intensity radio diffuse Galactic emission 
and cannot be applied to the polarized emission, since the 
two come from different spatial regions.}. 
The polarization angle map at the two frequencies 
should be the same in the absence of depolarization effects; we adopted the 
polarization angle map at 1420~MHz as intrinsic.
These four maps are the input skies for our 
toy-simulation of Faraday depolarization.
To build an $RM$ map we exploited the catalog produced by 
Spoelstra~(1984)\footnote{
The RM map obtained interpolating the values of Spoelstra~(1984)
are most likely not reliable. Those values were
obtained performing a linear fit to the five frequencies
data of the Leiden surveys and assuming as a result of the 
fit the lower possible RM.
Furthermore beamwidths effects were considered to be negligible.
Consequently the RMs derived by Spoelstra~(1984) should be
taken as indicative values.
We simply exploited the existence of these data to perform
toy-simulation of differential Faraday depolarization 
effects on the APSs of polarized emission. 
}
; it contains $\sim 1000$~~$RM$ data, 
whose minimum and maximum angular distance is  
 $d_{min} \sim 0.02^{\circ}$ and $d_{max} \sim 9^{\circ}$.
We interpolated these data to generate an $RM$ map with 
$\theta_{pixel}\simeq 1.84^{\circ}$~\footnote{The number of data points  
is about half the lowest one of the Leiden surveys catalogs, 
therefore we could not work with $\theta_{pixel}\simeq 0.92^{\circ}$.} 
(see Fig.~\ref{DepMap408}); 
the algorithm searches in circles of increasing radius from 
each pixel centre, considering radii from $d_{min}$ to $d_{max}$. 
It stops as soon as it finds at least 3 points and  
associates to the pixel the weighted average of 
the corresponding values; the weight is inversely proportional to
$d^{n}$, where $d$ is the distance from the pixel center. 
The results we show below have been obtained choosing $n=1$, 
although we checked that they remain  
valid also for $n = 0.5, 2, 4$.\\
We use the $RM$ map to transform the intrinsic $\phi$ and $PI$ for  
Faraday rotation and differential Faraday depolarization:
$$\phi_{output} \to \phi_{input} + \Delta\phi$$ 
$$PI_{output} \to PI_{input} \cdot |sin(\Delta\phi)/\Delta\phi|$$~.
At 1420~MHz the input and output maps are very similar.
At 408~MHz the situation is completely different: 
the large-scale structure appears significantly 
depressed in the depolarized map, whereas the small scale 
features become much more relevant 
(see Fig.~\ref{DepMap408}).

\begin{figure*}
\centering
\includegraphics[width=8cm,angle=90]{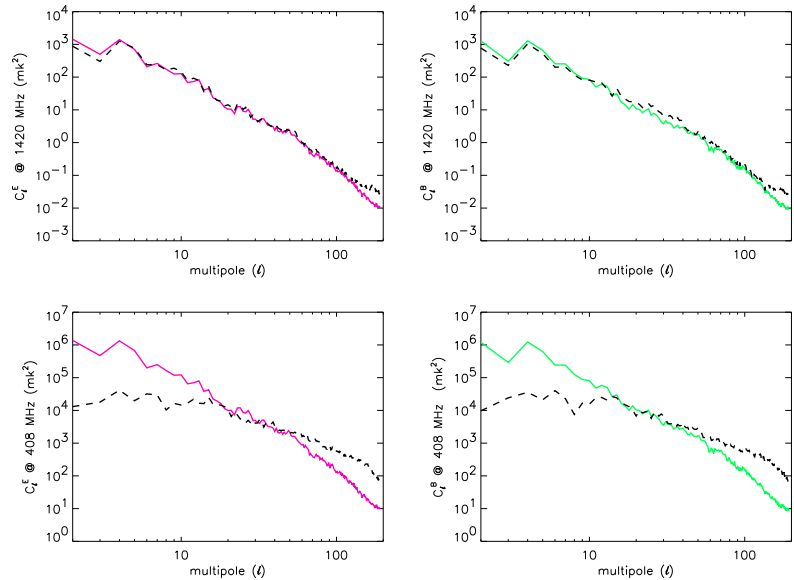}
\caption{ $E$ and $B$ modes APSs before (solid lines)
and after (dashed lines) the transformation according to Faraday depolarization. 
 The upper panels refer to 1420 MHz and
the lower ones to 408 MHz. The different importance of the 
depolarization effects at the two frequencies is evident: the 1420 MHz APSs are almost
unaffected, whereas the 408 MHz APSs significantly flatten.} 
\label{CompAPS_dep}
\end{figure*}

In Fig.~\ref{CompAPS_dep} we compare the APSs of the input and the output maps
at both the frequencies. At 1420~MHz the APSs are almost identical; on 
the contrary at 408~MHz the depolarized map APSs significantly flatten,
as expected from the increase of small scale features.
This supports the interpretation of the bulk of 
the APS flattening with decreasing frequency 
resulting from our analysis of the Leiden surveys 
in terms of the Faraday depolarization effect.

\section{Discussion and conclusions}

\begin{figure*}
\centering
\includegraphics[scale=0.8,angle=0]{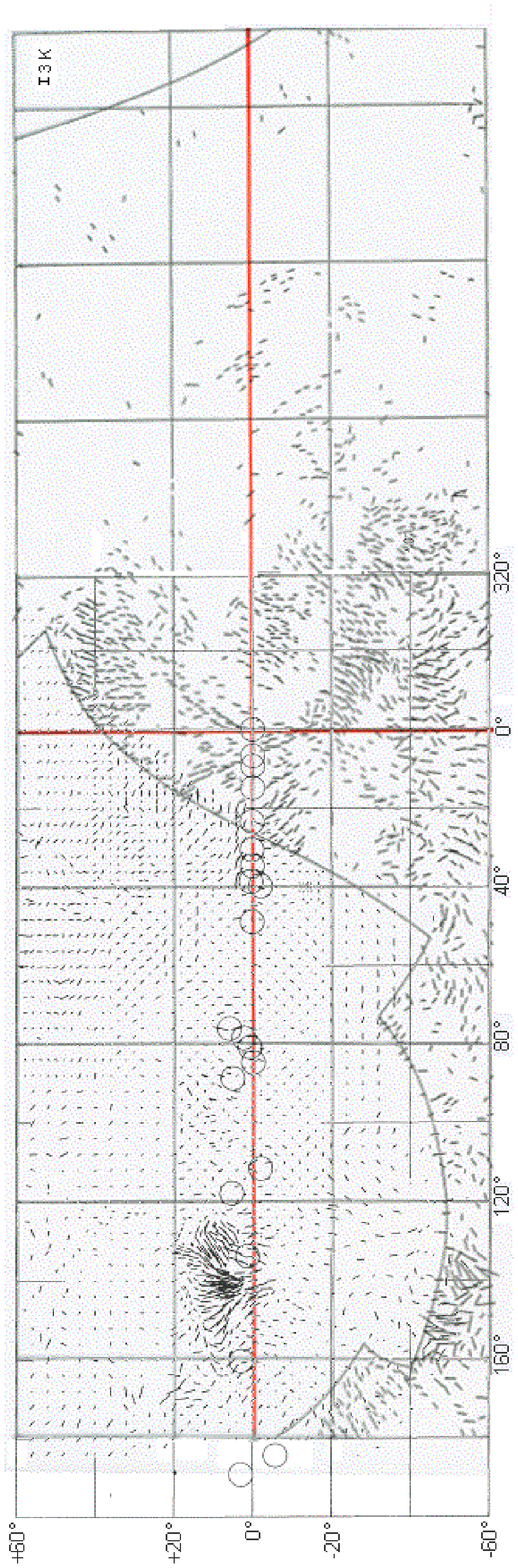}
\caption{Map of polarization vectors at 408 MHz, obtained combining
the corresponding figures of the Leiden survey (Brouw \& Spoelstra 1976)
and of the Parkes southern sky survey (\cite{mathewson65}).
Unfortunately, the vector length has a different scale in the two
original figures, therefore only a direct comparison of the 
direction is feasible.} 
\label{allpol408}
\end{figure*}

The Leiden surveys have been up to now the only
available data at $\nu \sim 1$~GHz
suitable for a multifrequency study of 
the polarized diffuse component of the Galactic synchrotron emission 
on large angular scales.
A linear polarization survey
of the Southern sky had previously been carried out at 408~MHz
using the Parkes telescope (\cite{mathewson65});
unfortunately these data have never been put in digital form.
In Fig.~\ref{allpol408} an image of the data from this survey has been 
combined with the Leiden survey at 408~MHz 
 to show an all-sky linear polarization map. 
 
The ``North Polar Spur'', the ``Fan Region'', and the ``Cetus Arc''  
(see \cite{berkhuijsen71} for a sketch of the Galactic Loops
these structures are thought to belong to) 
are visible as areas where the polarization vectors
are longer (the length is proportional to the polarized
intensity) and more regularly aligned, drawing small circles in the
sky. The most plausible theory suggests
that these Galactic spurs are the brighter ridges of 
evolved Supernova Remnant sitting relatively close to the Sun 
(within a few~$\times 10^2$~pc) and obscuring our
view of the Galactic diffuse non-thermal emission. 
The APSs obtained for these
regions will be characterized by an amplitude
considerably higher than the average, representing therefore
an upper limit for the polarized signal of the Galactic 
synchrotron radiation. On the other hand, 
the APS should be steeper in these regions, 
since the polarization vectors appear rather ordered and
with approximately constant length, implying a 
relatively low fluctuation power at small scales.
These considerations should be kept in mind
when extrapolating the results to the microwave 
range in order to check the impact of the 
Galactic synchrotron emission on CMB polarization 
measurements. \\
Implementing an interpolation method to deal with
the sparse and irregular sampling of the Leiden surveys, 
we produced maps of $PI$, $Q$ and $U$ 
with $\theta_{pixel} \simeq 0.92^{\circ}$
at five frequencies between 408 and 1411~MHz.
We accomplished several tests in order to check the reliability
of the interpolation algorithm and to optimize its quality
(see Sects.~3 and 4). 
We emphasize that the APSs computed from our maps
are significantly different from those obtained by other authors 
(e.g. \cite{bruscoli02}) for similar areas of the sky, using
the same data; namely, at the higher frequencies 
of the Leiden surveys we find steeper APSs
(first recognized in \cite{buriganalaporta02}). 
A firm confirmation of our approach comes  
from the comparison of the APSs obtained for the 1411~MHz map
with those of a preliminary version of the 
DRAO 1420~MHz survey, characterized by much denser sampling
and higher sensitivity.

We used the interpolated full coverage maps to study 
the APS of the polarized synchrotron emission
on the multipole range [2,50], being limited at high 
$\ell$ by the map sensitivity (related to the accuracy 
and sampling of the original data). 
Exploiting some better sampled regions, approximately
coinciding with the two most prominent Galactic spurs,
we could extend our analysis up to $\ell \sim 100$.
 
At each frequency the APSs of the entire coverage maps
 turn out to change slope at $\ell \sim 10$, 
 $C_{\ell}^{PI}$,  $C_{\ell}^{E}$,
and $C_{\ell}^{B}$ presenting a steepening 
going from the lower to the higher multipoles (see Sect.~5). 
A general and remarkable result (see Sect.~5) is that the 
slope of the polarized synchrotron emission APS increases 
from 408~MHz to 1411~MHz; it changes from $\alpha \sim -(1$-1.5)
to $\alpha \sim -(2$-3) for $\ell \sim 10$ to $\sim 100$
and from $\alpha \sim -0.7$ to $\alpha \sim -1.5$ for lower multipoles,
the exact value depending on the considered sky region
and polarization mode. \\
One possible explanation for such a behaviour of the APS
relies on Faraday depolarization arguments. 
Fluctuations of the electron density and/or of the 
magnetic field (in strength and/or direction) might redistribute the 
synchrotron emission power from the larger to the smaller angular 
scales, creating fake structures. 
The effect is more relevant at lower frequencies, where
the APS slope is expected to become flatter.
We have verified this interpretation
by performing a toy-simulation of pure Faraday depolarization
 in a simple {\it slab model}.  
We note that in the better sampled patches 
the polarization vectors are well
ordered and similarly distributed at 820 MHz and
1411 MHz, which is consistent with the slab model
assumptions at least for the considered angular scales.
We reproduced the effects of Faraday depolarization
on simulated polarized intensity maps, representing the intrinsic 
Galactic synchrotron emission at 408~MHz and 1420~MHz. 
We constructed an $RM$ map interpolating the data 
of Spoelstra~(1984) and applied it to transform the
input maps according to Faraday depolarization formulae. 
At 1420~MHz the input and output maps are almost identical, whereas
at 408~MHz the large-scale structure of the output map appears
depressed and the small scale structures more abundant.
Consequently, the corresponding 408~MHz APSs are flatter than the original ones 
(see Sect.~7.2 for a more precise description of the simulation process). 
It is therefore preferable to extrapolate the results
(spectral index and amplitude of the APS) obtained at 1411~MHz
(significantly less affected by Faraday depolarization effect
than those at 408~MHz) 
to estimate the synchrotron APS at microwave frequencies.

By interpreting the ratios of the APS amplitudes 
at 820 and 1411~MHz (the latter smoothed 
to the lower angular resolution of the former)
in terms of synchrotron emission depolarized 
by Faraday effect, we have identified possible $RM$ 
ranges 
(see Sect.~7.1). 
 This seems a reasonable approach at least in the case 
 of patch 1,~2,~and~3, since the polarization vectors 
 are mostly aligned, indicating rather homogeneous 
 physical conditions. 
Furthermore, taking into account the upper limit to $RM$
derived from the available information on 
the degree of polarization at 1.4~GHz
and considering that 
the maximum theoretical degree of synchrotron polarization is $\sim 
75\%$, we could break or, at least, reduce the degeneracy between the 
identified $RM$ intervals. 
For the three better sampled patches, 
the most reasonable value of $RM$ are $\sim 9-17$~rad/m$^2$.
However, given the uncertainty on the measured polarization
degree, $RM$ values in the interval $\sim 53-59$~rad/m$^2$
cannot be excluded.
Higher frequencies observations (at $\nu \gsim 2$-3~GHz) 
would be extremely useful to break the degeneracy 
left by the available $\sim 1$~GHz data.
\begin{acknowledgements}
We are grateful to W.~Reich for many constructive discussions
on the topic and for a careful reading of the original
manuscript. 
We aknowledge M.~Tucci and P.~Leahy  
and thank T.A.T. Spoelstra, R.~Wielebinski and R.~Beck. 
L.L.P. thanks E.~Hivon.
C.B. thanks W.~Reich, R.~Wielebinski,
and the MPIfR, Bonn for the warm hospitality.
L.L.P. was supported for this research through a stipend
from the International Max Planck Research School (IMPRS)
for Radio and Infrared Astronomy at the Universities of
Bonn and Cologne.
Some of the results in this paper have been derived using 
the HEALPix package (G\'orski et al.~2005).
We warmly thank the anonymous referee for useful comments.

\end{acknowledgements}

\end{document}